%% file: main.tex
\documentclass[a4paper,11pt]{article}
\usepackage[utf8]{inputenc}
\usepackage{textgreek}

\usepackage{jcappub} 
\usepackage[T1]{fontenc} 
\usepackage{array}
\title{\boldmath CMB Hemispherical Power Asymmetry from Early Phase of Inflation}
\usepackage{gensymb}
\usepackage{hyperref}
\usepackage{booktabs}      
\usepackage{graphicx}      
\usepackage[utf8]{inputenc}
\usepackage{textcomp}
\usepackage{amsmath}
\usepackage{caption}
\usepackage{comment}
\usepackage{subfigure}
\usepackage{subcaption}
\input{def}

\author[a]{Akash Gandhi,}
\author[a]{Mohit Panwar}
\author[a,b]{and Pankaj Jain}

\affiliation[a]{Department of Physics, Indian Institute of Technology, Kanpur 208016, India}
\affiliation[b]{Department of Space, Planetary \& Astronomical Sciences \& Engineering (SPASE), Indian Institute of Technology, Kanpur 208016, India

}

\emailAdd{gakash@iitk.ac.in}
\emailAdd{panwarmohit706@gmail.com}
\emailAdd{pkjain@iitk.ac.in}

\abstract{
We investigate the hemispherical power asymmetry observed in the CMBR by attributing it to an early inhomogeneous phase of cosmic expansion. Unlike the conventional assumption of a perfectly isotropic and homogeneous pre-inflationary Universe, we introduce a small inhomogeneous perturbation, treated within a perturbative framework. Our analysis builds on previously developed empirical models of inhomogeneous primordial power spectrum models based on dipole modulation. Using in-in formalism, we compute two-point correlations directly from the metric and demonstrate that, at leading order, this introduces a direction-dependent power spectrum that breaks rotational symmetry and naturally selects a preferred direction, relating observed violation of the cosmological principle to inflationary power spectra arising from scalar field fluctuations. Additionally, we find that this framework produces correlations between multipoles separated by $\Delta l=1$, leading to distinctive signatures in the multipole space. Furthermore, we constrain the parameters of the inhomogeneous perturbation using observed PR4 \texttt{Commander} CMB data.}


\begin{document}
\maketitle
\flushbottom
\section{Introduction}
The standard $\Lambda$CDM cosmological model is built upon the cosmological principle, which assumes that the universe is homogeneous and isotropic on large scales. This assumption is supported by observations of the cosmic microwave background (CMB), which exhibits near-uniform temperature fluctuations across the sky. However, various studies using data from WMAP and Planck have revealed anomalies suggesting deviations from statistical isotropy, leading to questions about the validity of this foundational principle. One of the most prominent deviations is the hemispherical power asymmetry in the CMB \cite{2004ApJ...605...14E,refId0,Zibin:2015ccn,Planck:2015igc,Akrami:2014eta}, where one half of the sky shows a slightly higher fluctuation amplitude than the other. This unexpected variation challenges the assumption that primordial density perturbations should be directionally uniform. Other interesting anomalies include dipole in radio polarizations \citep{Jain1998} and alignment of the quadrupole and octopole moments \cite{2004PhRvD..69f3516D,Gordon_2005} of the CMB, both of which point close to the CMB dipole \citep{Ralston2004, Land_2005}. Additionally, large-scale galaxy surveys \cite{Blake_2002,2011ApJ...742L..23S,2012MNRAS.427.1994G,2013A&A...555A.117R,Secrest_2021} and X-ray clusters \cite{2010ApJ...712L..81K} also suggest a large scale anisotropy aligned approximately with the CMB dipole. We point out that so far, the observed deviations from isotropy are only suggestive and require more data for confirmation.

There have been many theoretical frameworks that have been proposed to explain potential isotropy violations. The ones based on non-commutative geometry \cite{E_Akofor_2008, Jain_2015, PhysRevD.94.063531} or loop quantum cosmology \cite{Agullo:2021oqk} predict specific forms of anisotropy that could manifest in CMB observations.
Bianchi Type I Models \cite{10.1093/mnras/stu932} can address some large-scale CMB anomalies, they produce only quadrupolar anisotropies and no B-mode polarization. Bianchi Type VIIh Models \cite{10.1111/j.1365-2966.2007.12221.x} predict significant B-mode polarization (comparable to E-mode power) and parity-violating correlations that are not observed in current data, casting doubt on their viability. The kinetic and thermal SZ effects \cite{refId2} from local large-scale structure can contribute to apparent isotropy violations, though recent studies suggest they cannot fully account for observed anomalies \cite{refId2, PhysRevD.101.123508}. The proposal that we are located in a local underdensity can explain some isotropy violations \cite{refId3} through the lensing and integrated Sachs-Wolfe effects. It is suggested in \cite{Nistane_2019} that we might be located near the center of a large void, which could create apparent isotropy violations. The CMB sky measured by an off-center observer in such voids would not be statistically isotropic and would exhibit lensing-like distortions. While curvaton models \cite{PhysRevD.90.023523, Lyth:2006gd} can produce local-type non-Gaussianity, their predictions for scale-invariant asymmetry and non-Gaussianity are not strongly favored by current observations.

Inhomogeneous Models, including Lemaitre-Tolman-Bondi Models \cite{Bolejko_2011}, can potentially explain apparent cosmic acceleration without dark energy but require fine-tuning of the observer's location and face constraints from structure formation observations. The Szekeres Models \cite{PhysRevLett.111.251302}, which are a class of inhomogeneous models, can fit supernova data as well as $\Lambda$CDM, they fail to reproduce the observed late-time suppression in structure growth unless a cosmological constant is included.

Most theoretical models that address CMB anomalies face the challenge of explaining multiple seemingly unrelated features simultaneously without producing other observational consequences that are ruled out by data. A potential explanation for these deviations involves the presence of superhorizon perturbations, which are fluctuations with wavelengths larger than the observable universe \cite{PhysRevD.72.083501}. These perturbations could imprint a preferred direction on the universe \cite{Erickcek_2008}, leading to the observed anisotropies. Models incorporating superhorizon scalar fields within the curvaton model and phenomenological inflationary mechanisms that modulate small-scale power—such as isocurvature perturbations \cite{PhysRevD.80.083507, PhysRevD.42.313} and isocurvature-driven power gradients \cite{JohnMcDonald_2013, Assadullahi_2015} have been proposed. These models \cite{PhysRevD.78.123520} can yield the required scale-dependent power asymmetry while remaining consistent with small-scale isotropy.

The observed deviations from isotropy suggest a violation of the cosmological principle, and hence appear to suggest that the background space-time metric may be different from the standard FRW metric. It is important to realize that within the Big Bang paradigm, the background metric need not be FRW at all times. Indeed, it is postulated that at very early times, the metric may not obey the cosmological principle and acquires isotropy and homogeneity during inflation \citep{Wald_Robert_1983}. This raises the possibility that the perturbation modes during the very early phase of inflation may not obey statistical isotropy and homogeneity. It has been suggested that for a certain range of parameters, these modes may affect cosmological observations today at large distance scales \cite{PhysRevD.75.083502,Rath:2013bfa,Rath:2014cka}. Hence, these may provide an explanation for the observed large-scale anisotropies. 

It is crucial to distinguish between models that break rotational invariance while maintaining homogeneity from models that break homogeneity. A seminal example of the former is the work by Ackerman et al.~\cite{PhysRevD.75.083502}, which investigates the breakdown of rotational invariance in the early universe using the in-in formalism \cite{Weinberg:2005vy}. These homogeneous but anisotropic models successfully predict quadrupolar features in the CMB power spectrum, making them well-suited for explaining the alignment of the quadrupole and octopole moments \cite{Rath:2013bfa}. However, as demonstrated in earlier analyses~\cite{Jain_2015,PhysRevD.94.063531}, a homogeneous anisotropic universe fundamentally fails to generate the odd-parity correlations required to explain the Hemispherical Power Asymmetry (HPA) unless the space-time is non-commutative. Specifically, such models do not produce the dipolar modulation signature that correlates multipoles $\ell$ and $\ell \pm 1$.

This work aims to determine whether an early inhomogeneous phase can generate a hemispherical power asymmetry in the CMB using the in-in formalism \cite{Weinberg:2005vy}. We find that such a phase can lead to dipole modulation and, hence, hemispherical anistropy. Furthermore, we make a detailed comparison of the theoretical predictions with the observed multipole dependence of the power.


\section{Hemispherical Power Asymmetry}
The hemispherical power asymmetry in CMB has been detected in both WMAP 9-year ILC and Planck datasets. Analysis of the full-sky temperature map reveals a power asymmetry, i.e., the northern hemisphere in the ecliptic coordinate system exhibits lower variance compared to the southern hemisphere.
The statistical significance of this effect has been found at the level of 3$\sigma$–3.5$\sigma$ 
\cite{refId0,Zibin:2015ccn,Planck:2015igc,Akrami:2014eta}.

After the initial detection in WMAP temperature maps, hemispherical power asymmetry (HPA) has been re-examined across subsequent releases using dipole-modulation fits, hemispherical power ratios, Bayesian statistics, and multi-scale estimators \cite{Eriksen_2007, 10.1111/j.1365-2966.2007.12707.x, Lew_2008, PhysRevD.78.103504, Groeneboom2010ImprintsOA, 10.1093/mnras/stt1219, Hansen_2009, 10.1111/j.1365-2966.2010.16905.x, SamuelFlender_2013, Freeman_2006}.  Planck-era analyses revisited its significance, including local-variance/LVE and polarization extensions, and evaluated power-asymmetry significance across scales with improved masking and systematics controls \cite{sanyal2025reassessmentlvemethodhemispherical, Gimeno-Amo_2023, Quartin_2015, aluri2017powerasymmetrycmbpolarization}. These studies confirm
a hemispherical power asymmetry at the $\simeq 3\sigma$ level when the analysis is restricted to large angular scales ($\ell \lesssim 60$), in good agreement with the WMAP-based estimates. At the same time, they show that the best-fit modulation amplitude $A(\ell)$ decreases with multipole and becomes consistent with zero for $\ell \gtrsim 600$, so that the overall significance of the effect diminishes when the full range of angular scales is included. Kinematic effects from aberration and Doppler boosting were quantified as an additional (but distinct) source of hemispherical modulation \cite{Notari_2014}.

Several theoretical mechanisms, in addition to the superhorizon modulation scenario of \cite{PhysRevD.78.123520}, have been proposed to explain the observed hemispherical power asymmetry. One early phenomenological explanation is the linear modulation model of \cite{Gordon_2007}, in which the primordial perturbations are modulated by a dipolar spatial dependence. More explicit realizations based on an inhomogeneous primordial power spectrum were developed in \cite{Rath:2014cka, Kothari:2015tqa}. In the model of \cite{Rath:2014cka}, the power spectrum depends weakly on spatial position and leads to correlations between CMB multipoles separated by $\Delta\ell=1$, thereby providing a characteristic off-diagonal signature in harmonic space. In contrast, the model discussed in \cite{Kothari:2015tqa} emphasizes large-scale damping effects from an inhomogeneous or anisotropic primordial spectrum and does not predict the same $\ell$ to $\ell+1$ correlation pattern as a generic outcome. In both cases, the associated model parameters have been constrained by comparison with CMB observations.

Complementary proposals include modulated reheating and space-dependent adiabatic components that also affect the scalar spectral index $n_s$ as well as scenarios in which the hemispherical power asymmetry is correlated with a spatial modulation of the tensor-to-scalar ratio \(r\) and the primordial tensor (B-mode) signal \cite{John_McDonald_2013, PhysRevD.89.127303, McDonald_2014, 10.1093/mnras/stu921}. Consistency conditions relate the dipole amplitude to squeezed-limit non-Gaussianity, motivating scenarios with scale-dependent $f_{\rm NL}$ or more general initial states \cite{PhysRevD.88.083527, Namjoo_2014, Firouzjahi_2014, Byrnes_2015, PhysRevLett.110.011301}. 

A comprehensive analysis of such scenarios was performed by Byrnes et al.~\cite{PhysRevD.93.123003}. They demonstrated that while these models can phenomenologically reproduce a power asymmetry, they face severe theoretical constraints. In curvaton-type models, the asymmetry amplitude $A$ typically scales with the local non-Gaussianity parameter $f_{NL}$ and the fractional perturbation of the curvaton field $\delta \sigma_L / \sigma$, following the relation $A \sim f_{NL} \delta \sigma_L / \sigma$. Consequently, generating the observed asymmetry amplitude of $A \approx 0.07$ generally requires a large primordial bispectrum, $|f_{NL}| \gtrsim 60$. This requirement is in strong tension with the stringent constraints from Planck, which find $f_{NL}^{\text{local}}$ consistent with zero ($f_{NL}^{\text{local}} = 0.8 \pm 5.0$). To evade this, such models often require extreme fine-tuning of initial conditions or scale-dependent bispectra.

Beyond these mechanisms, a broader space of models has been explored, including modulation sourced by an anisotropic stochastic gravitational-wave background and unified scalar–tensor origins, with polarization/B-mode tests as key discriminants \cite{mukherjee2016unifiedoriginhemisphericalasymmetry, PhysRevD.91.062002, PhysRevLett.116.221301}.  Direction-dependent parameter inference under HPA has been investigated with non-isotropic simulations \cite{Mukherjee_2016}, and bispectrum/response-function approaches show how the asymmetry can be generated while respecting bispectrum constraints in tuned cases \cite{Byrnes_2016}.  Early-universe/UV completions and alternatives include primordial domain walls, pre-inflationary topological defects, large $g_{\rm NL}$ modulation, loop-quantum-cosmology/quantum-gravity effects, sound-speed modulation, and general forecasts that motivate closing in with polarization and multi-probe constraints \cite{Jazayeri_2014, PhysRevD.96.083516, PhysRevD.92.023505, PhysRevD.92.064038, Wang_2016, PhysRevD.93.123003, PhysRevD.97.043501, PhysRevD.97.063504, PhysRevD.87.123005, PhysRevD.89.023005}.

Anisotropic background cosmologies such as Bianchi type~VII$_h$ generate deterministic spiral-like CMB templates, incorporating a dark-energy component that constrains these models tightly and disfavor them as a full explanation of this observed anomaly \cite{Jaffe_2006}. 
Alternatively, a homogeneous primordial magnetic field explicitly breaks spatial isotropy and induces characteristic off-diagonal CMB correlations that can mimic a dipole-like modulation \cite{PhysRevD.78.063012}.

At the phenomenological level, the effect of hemispherical asymmetry on the temperature field is commonly modeled by a dipole modulation of an underlying statistically isotropic sky.\cite{Gordon_2005,Kothari:2015tqa, Bennett:2010jb, Axelsson:2013mva,PhysRevD.71.083508, Pranati_K_Rath_2013}
In the low-$\ell$ regime, this is described by
\begin{equation} \label{eq:temp_modulation}
    \Delta T_{\rm obs}(\hat n)=\Delta T_{\rm iso}(\hat n)\left(1+A\hat\lambda\cdot\hat n\right)
\end{equation}
where $\Delta T_{\rm obs}(\hat n)$ is the temperature fluctuation observed in the direction $\hat n$, $\Delta T_{\rm iso}(\hat n)$ is a statistically isotropic field and $A$ is the amplitude of the dipole modulation.
This also leads to correlations between multipoles $\ell$ and $\ell \pm 1$ which have been detected in data \cite{Pranati_K_Rath_2013}.
Moreover, the inferred modulation amplitude is scale-dependent: it is largest at low multipoles and decreases towards higher $\ell$ \cite{Bennett:2010jb,Ghosh:2015qta,Adhikari:2014mua,Planck:2015igc}. Harmonic decomposition of the masked sky estimates the asymmetry’s significance at approximately 3$\sigma$, with the asymmetry axis pointing toward $(224 \degree, -22 \degree)$ in galactic coordinates.
The statistical significance decreases when considering the full range of scales, and the effect is dominant at large distance scales, suggesting a possible connection to the early phase of inflation \cite{Akrami:2014eta,Koivisto:2010fk}.


\section{Power Spectra due to Inhomogeneous Early Phase of Inflation}

In the present paper, we explain the hemispherical anisotropy in terms of an early inhomogeneous phase of inflation. We assume that at early times, Universe does not follow the cosmological principle \cite{Wald_Robert_1983} and the metric deviates from the standard FRW metric. The deviation is assumed to be small and can be treated perturbatively.
We consider the following simple model for the background inhomogeneous metric,
\begin{equation}\label{eq:background_metric}
    ds^2=-(1+2 \Psi) d t^2 + a^2(t)(1-2 \Phi)\delta_{ij}dx^{i}dx^{j}
\end{equation}
 with  
\begin{equation}\label{eq:superhorizon_mode}
\Psi =\alpha \sin \left(\kappa z+\omega\right)\,.
\end{equation} 
We set the two scalar potentials equal, $\Phi = \Psi$. The assumption $\Phi = \Psi$ is made to simplify the analysis. In General Relativity, this equality holds in the absence of anisotropic stress sources~\cite{PhysRevD.77.123541}. In general, this need not apply, but deviations from this will introduce more parameters which we wish to avoid here and may be postponed to future work. We assume $\Psi$ acts as a small perturbation. 
We expect that the inhomogeneities would be prominent only during the early phase of inflation and would decay with time. This can be implemented by introducing a suitable decay factor with the perturbation. In our case, we will apply this model only to large-distance observables, and this decay factor makes no difference in our final result. Even if we consider decaying amplitude of sinusoidal perturbation mode, such as, $\alpha_0 e^{-\gamma t}$, it is easy to show that g(p) does not change significantly for small $p$. Hence, the factor does not affect the structure of the two‑point correlation function for large wavelength modes. At this stage, we do not know what the full metric might be and, hence, choose a simple model. In order to go beyond this we will need to make an elaborate model along with proper matter content and systematically evolve the metric to the final FRW form. This will require considerable more effort and has been postponed to future work. We point out that several Bianchi models show this behavior \cite{Wald_Robert_1983} and we are working in analogy with these models. Essentially, we assume that, as in the case of Bianchi models, the deviation from isotropy and homogeneity decays quickly during inflation.

\subsection{Two point Correlations}
We proceed with the investigation of the early inhomogeneous inflationary phase by introducing a small spatial metric perturbation on a flat FLRW background. Specifically, we assume a metric of the form of Eq. \eqref{eq:background_metric} with Eq. \eqref{eq:superhorizon_mode} representing a slight sinusoidal inhomogeneity along one spatial direction. We then evolve a canonical scalar inflaton field on this perturbed background using the in-in (Schwinger–Keldysh) formalism \cite{Weinberg:2005vy} to compute primordial two-point correlations.

We start from a single canonical scalar field $\phi$ with a potential $V(\phi)$, which is minimally coupled to gravity and standard model for single-field slow-roll inflation. The resulting dynamics is governed by the action,
\begin{equation}\label{eq:action}
    S[g_{\mu\nu},\phi] = \int d^4x \sqrt{-g} \left\{ \frac{1}{16\pi G}R - \frac{1}{2}g^{\mu\nu} \partial_\mu \phi \, \partial_\nu \phi - V(\phi) \right\}
\end{equation}

From this action, the Lagrangian density corresponding to the inflationary scalar field $\phi$ is given by $\mathcal{L}=\sqrt{-g}[-\frac{1}{2}g^{\mu\nu}\partial_{\mu}\phi\partial_{\nu}\phi-V(\phi)]$. The minus sign of the kinetic terms comes from the fact that we are working with metric signature $-+++$.  The equation of motion of the field $\phi$ can be obtained formally by using 
\begin{equation}
    \frac{1}{\sqrt{-g}}\, \partial_{\mu}\!\bigl(\sqrt{-g}\,g^{\mu\nu}\,\partial_{\nu}\phi\bigr) = V_{,\phi}
\end{equation}
where $V_{,\phi} = \partial V/\partial\phi$. As explained earlier, we set  $\Phi=\Psi$ for the remainder of this derivation. 

We are interested in obtaining the interaction Hamiltonian due to the metric \ref{eq:background_metric}. Using $\sqrt{-g} = a^3 (1 - 2\Psi) + \mathcal{O}(\Psi^2)$ and expanding about $\phi(t)$, using the background equation
$\ddot\phi+3H\dot\phi+V_{,\phi}=0$ to remove linear terms in $\delta\phi$, and
dropping total derivatives, the quadratic Lagrangian for $\delta\phi$ up to
$\mathcal O(\Psi)$ is
\begin{equation}
\mathcal L_2[\delta\phi]
= \frac12 a^{3}(1-4\Psi)\,\dot{\delta\phi}^{\,2}
  - \frac12 a\,(\nabla\delta\phi)^2
  - \frac12 a^{3}(1-2\Psi)\,V_{,\phi\phi}\,\delta\phi^2.
\label{eq:L2}
\end{equation}
Notice that no linear $\Psi$ correction survives in the gradient sector, since
$a^3(1-2\Psi)\times (1+2\Psi)/a^2 = a + \mathcal O(\Psi^2)$.
The conjugate momentum and its linearized inversion read
\begin{equation}
\delta\pi \equiv \frac{\partial\mathcal L_2}{\partial \dot{\delta\phi}}
= a^{3}(1-4\Psi)\,\dot{\delta\phi},
\qquad
\dot{\delta\phi}
= \frac{1+4\Psi}{a^{3}}\,\delta\pi \;+\; \mathcal O(\Psi^2).
\label{eq:momenta}
\end{equation}
Performing the Legendre transform, the quadratic Hamiltonian density becomes
\begin{equation}
\mathcal H_2
= \frac{1}{2a^{3}}(1+4\Psi)\,\delta\pi^{2}
  + \frac12 a\,(\nabla\delta\phi)^2
  + \frac12 a^{3}(1-2\Psi)\,V_{,\phi\phi}\,\delta\phi^2.
\label{eq:H2}
\end{equation}
We split $\mathcal H_2=\mathcal H_0+\mathcal H_{\rm int}$ with
\begin{align}
\mathcal H_0
&= \frac{1}{2a^{3}}\,\delta\pi^{2}
 + \frac12 a\,(\nabla\delta\phi)^2
 + \frac12 a^{3} V_{,\phi\phi}\,\delta\phi^2 ,
\label{eq:H0} \\[2mm]
\mathcal H_{\rm int}
&= \frac{2\Psi}{a^{3}}\,\delta\pi^{2}
 - a^{3}\Psi\,V_{,\phi\phi}\,\delta\phi^2
 = 2a^{3}\Psi\,\dot{\delta\phi}^{\,2}
 - a^{3}\Psi\,V_{,\phi\phi}\,\delta\phi^2 .
\label{eq:Hint-density}
\end{align}
Thus the (instantaneous) interaction Hamiltonian is
\begin{equation}
H_{\rm int}(t)=\int\!\mathrm{d}^3x\;
\Big[\,2a^{3}\Psi\,\dot{\delta\phi}^{\,2}
      - a^{3}\Psi\,V_{,\phi\phi}\,\delta\phi^{2}\Big].
\label{eq:Hint}
\end{equation}    
Using the definition of slow-roll parameter 
\begin{equation}
\eta_V \equiv M_{\mathrm{Pl}}^2 \frac{V_{,\phi\phi}}{V}
\end{equation}
we can write the additional term $- a^3 \Psi V_{,\phi\phi}(\phi) \delta\phi^2$ = $- 3a^3 \Psi H^2 \eta_V  \delta\phi^2$. As $\eta_V \ll 1$ during slow roll, the potential contribution is suppressed relative to the kinetic term by a factor of $\eta_V$, justifying its neglect within the slow-roll framework. This derivation provides the correct, physically meaningful interaction Hamiltonian for a canonical scalar field in a perturbed FLRW universe. It is built upon the standard action and proceeds via a gauge-invariant formalism, ensuring the results are physical.

It is important to clarify the validity of using standard quantization procedures in this inhomogeneous setting. We employ the interaction picture, where the total Hamiltonian is partitioned into a free part $H_0$, governing the unperturbed homogeneous background, and an interaction part $H_{\text{int}}$ that encapsulates the metric inhomogeneity $\Psi$. Because $\Psi$ is treated as a small perturbation (we keep terms first order in $\alpha$), the
leading effect of the inhomogeneity enters through $H_{int}$. In the Interaction Picture, the field operators $\delta \phi(t, \vec{x})$ evolve according to the free equations of motion derived from $H_0$. Since $H_0$ is defined on a homogeneous background, the mode functions $u_k(\eta)$ used to expand the operators are the standard solutions (Hankel functions) for a massless scalar in de Sitter space. The inhomogeneity is not ignored, the spatial dependence of $\Psi(\vec{x})$ enters through the time integration of $H_{int}$. When calculating the correlation functions, the vertex factors from $H_{int}$ (which contain terms like $\int d^3x e^{i\vec{q}\cdot\vec{x}} \Psi(\vec{x})$) couple the different Fourier modes. The unperturbed evolution is governed by $H_0$, for
which the usual mode functions of a massless scalar in de Sitter space are appropriate. We therefore
quantise the scalar field in the Bunch–Davies vacuum and use the homogeneous mode functions in the free
theory.

The effectively massless scalar field fluctuation $\delta\phi(\vec{x}, t)$ is conventionally quantised in terms of Fourier modes as
\begin{equation}\label{eq:mode_expansion}
    \delta\phi(\vec{x}, t)=\int \frac{d^3 k}{(2 \pi)^3}\left(e^{i \vec{k} \cdot \vec{x}} \phi_k(t) a_k+e^{-i \vec{k} \cdot \vec{x}} \phi_k^*(t) a_k^{\dagger}\right)
\end{equation}
where the standard commutation relation $\left[a_k, a_{k^{\prime}}^{\dagger}\right]=(2 \pi)^3 \delta\left(\vec{k}-\vec{k}^{\prime}\right)$ between creation and annihilation operators holds.
The two-point correlations of the scalar field, up to first order, are given by \citep{Weinberg:2005vy}, 
\begin{equation}\label{eq:TPC_Weinberg}
\left\langle\phi\left(\vec{x_1}, t\right) \phi\left(\vec{x_2}, t\right)\right\rangle \equiv \left\langle\delta\phi\left(\vec{x_1}, t\right) \delta\phi\left(\vec{x_2}, t\right)\right\rangle_{iso}+ i \int_0^t d t^{\prime}\left\langle\left[H_{\rm int}\left(t^{\prime}\right), \delta\phi\left(\vec{x_1}, t\right) \delta\phi\left(\vec{x_2}, t\right)\right]\right\rangle
\end{equation}

During the evaluation of the two-point correlation functions using the in-in formalism in \ref{eq:TPC_Weinberg}, we kept the terms up to first order only retaining terms linear in $\Psi$ and discarding those quadratic or higher (which are of order $\mathcal{O}(\alpha^2)$). The solution to the unperturbed Euler-Lagrange equation of motion is given by
\begin{equation}\label{eq:mode_function}
    \phi_p^{(0)}(\eta)=\frac{H}{\sqrt{2 p}}\left(\frac{i}{p}-\eta\right) \exp (-i p \eta)
\end{equation}
These are positive-frequency solutions in the Bunch-Davies vacuum, the default vacuum choice in inflationary cosmology \cite{Maldacena2013}.
We evaluate the first term of Eq. \ref{eq:TPC_Weinberg}, using Eq. \ref{eq:mode_expansion}, as,
\begin{equation}\label{eq:TPC_isotropic}
\left\langle\delta\phi\left(\vec{x_1}, t\right) \delta\phi\left(\vec{x_2}, t\right)\right\rangle_{iso}=\int \frac{d^3 p}{(2 \pi)^3} e^{i \vec{p} \cdot\left(\vec{x}_1 - \vec{x}_2\right)}P_{i s o}(p)
\end{equation}
where $P_{i s o}(p) \simeq\left|\phi_p^{(0)}(\eta)\right|^2$. Here we take $p\approx a_I H$ and in the limit $\left|p\eta\right| << 1$\citep{Rath:2013bfa}, we find that
\begin{equation}\label{eq:spectrum_isotropic}
P_{i s o}(p) \simeq\left|\phi_p^{(0)}(\eta)\right|^2\simeq\frac{H^2}{2p^3}
\end{equation}
Now moving on to the second term in the Eq. \ref{eq:TPC_Weinberg}, we first evaluate
\begin{equation}
    \begin{split}
        \left\langle \left[ H_{\rm int}(t'), \delta\phi(\vec{x_1}, t)\, \delta\phi(\vec{x_2}, t) \right] \right\rangle = 
        2 a^3(t') \int d^3x\, \Psi \int \frac{d^3p\, d^3q}{(2\pi)^6} \Big[ & \\
        e^{i (\vec{p} + \vec{q}) \cdot \vec{x}}\, \dot{\phi}_p(t')\, \dot{\phi}_q(t')\, \phi_p^*(t)\, \phi_q^*(t) \, 
        \big( e^{-i \vec{q} \cdot \vec{x}_1 - i \vec{p} \cdot \vec{x}_2} + e^{-i \vec{p} \cdot \vec{x}_1 - i \vec{q} \cdot \vec{x}_2} \big) 
        & \\
        + e^{-i (\vec{p} + \vec{q}) \cdot \vec{x}}\, \dot{\phi}_p^*(t')\, \dot{\phi}_q^*(t')\, \phi_p(t)\, \phi_q(t) \, 
        \big( e^{i \vec{p} \cdot \vec{x}_1 + i \vec{q} \cdot \vec{x}_2} + e^{i \vec{q} \cdot \vec{x}_1 + i \vec{p} \cdot \vec{x}_2} \big)
        \Big]
    \end{split}
\end{equation}
Feeding in $\Psi$ from Eq. \ref{eq:superhorizon_mode} we obtain the integral,
\begin{equation}
\int e^{i (\vec{p}+\vec{q}) \cdot \vec{x}} \sin(\kappa z + \omega) \, d^3 x
= \frac{(2\pi)^3}{2i} \left[
e^{i\omega} \delta^{(3)}(\vec{p}\,' + \vec{q})
- e^{-i\omega} \delta^{(3)}(\vec{p}\,'' - \vec{q})
\right]
\end{equation}
where the magnitudes of $\vec{p}, \vec{p}\,'$ and $\vec{p}$\,'', 
respectively, are
\begin{equation}
p=\sqrt{p_x^2+p_y^2+p_z^2}
\end{equation}
\begin{equation}
    p^{\prime}=\sqrt{p_x^2+p_y^2+\left(p_z+\kappa\right)^2}
\end{equation}
\begin{equation}
    p^{\prime\prime}=\sqrt{p_x^2+p_y^2+\left(p_z-\kappa\right)^2}
\end{equation}

The final form of Eq. \ref{eq:TPC_Weinberg} becomes
\begin{equation}
\langle\phi(\vec{x_1},t)\phi(\vec{x_2},t) \rangle = \int \frac{d^3 p}{(2\pi)^3} e^{-i \vec{p} \cdot (\vec{x}_1 - \vec{x}_2)} \times 
\left[ P_{iso}(p) + (\hat{k} \cdot \vec{X}) g(p) \right]
\end{equation}
where \(\vec{X} = (\vec{x}_1 + \vec{x}_2)/2\) and 
\begin{equation}
\begin{split}
g(p) \simeq 2i\alpha\kappa \int_{-\frac{1}{Ha_I}}^{\eta} d\eta'\,
\left( -\frac{1}{H \eta'} \right)^4 \Big[
& e^{i \omega} \dot{\phi}_p^{(0)}(\eta') \dot{\phi}_{p'}^{(0)}(\eta') \phi_p^{*(0)}(\eta) \phi_{p'}^{*(0)}(\eta) \\
& + e^{-i \omega} \dot{\phi}_p^{*(0)}(\eta') \dot{\phi}_{p''}^{*(0)}(\eta') \phi_p^{(0)}(\eta) \phi_{p''}^{(0)}(\eta) \\
& + e^{i \omega} \dot{\phi}_p^{*(0)}(\eta') \dot{\phi}_{p'}^{*(0)}(\eta') \phi_p^{(0)}(\eta) \phi_{p'}^{(0)}(\eta) \\
& + e^{-i \omega} \dot{\phi}_p^{(0)}(\eta') \dot{\phi}_{p''}^{(0)}(\eta') \phi_p^{*(0)}(\eta) \phi_{p''}^{*(0)}(\eta)
\Big]
\end{split}
\end{equation}
Here we have kept terms up to leading order in $\kappa$ and have used,
\begin{equation}
\eta=\int \frac{d t}{a(t)}=-\frac{1}{H a_I} e^{-H t}
\end{equation}
In the limit $\left|p\eta\right| << 1$, the above expression is evaluated as
\begin{equation}
g(p)=\frac{\alpha H^2 \kappa}{2p} \left[\frac{e^{i\omega}}{p^{\prime}(p+p')} [1-\cos\left(\frac{p+p^{\prime}}{H a_I}\right)]+\frac{e^{-i\omega}}{p^{\prime \prime}(p+p'')} [1-\cos \left(\frac{p+p^{\prime \prime}}{H a_I}\right)]\right]
\end{equation}
At leading order, the presence of the sinusoidal scalar mode breaks statistical isotropy of the power spectrum, yielding a direction-dependent component. This violation of rotational symmetry introduces correlations between spherical harmonic multipoles separated by $\Delta \ell = 1$, a signature well-established in the literature for models involving dipole modulation \cite{PhysRevD.71.083508, PhysRevD.80.063004, Pranati_K_Rath_2013}. This means that, unlike the standard isotropic case (where $\langle a_{\ell m} a_{\ell' m'} \rangle \propto C_\ell\, \delta_{\ell \ell'}\, \delta_{m m'}$), the covariance matrix acquires off-diagonal elements for $\ell'=\ell\pm1$. Physically, this pattern corresponds to a dipolar modulation imprint on the primordial fluctuations.
\section{Linking Correlations with HPA}
The temperature fluctuations can be decomposed in terms of spherical harmonics
\begin{equation}\label{eq:Temp_spherical_harmonics}
    \frac{\Delta T(\hat{n})}{T_0}=\sum_{\ell=1}^{\infty} \sum_{m=-\ell}^\ell a_{\ell m} Y_{\ell m}(\hat{n})\,,
\end{equation}
where $T_{0} = (4\pi)^{-1}\int_{4 \pi} T(\hat{n})d\Omega$. These fluctuations can be related to the primordial density fluctuations as \cite{Dodelson:2020bqr}
\begin{equation}\label{eq:Temp_fluctuation_spectrum}
\frac{\Delta T(\hat{n})}{T_0}=\int d^3 p \sum_\ell \frac{2\ell+1}{4 \pi}(-i)^\ell P_\ell(\hat{p} \cdot \hat{n}) \delta(p) \Theta_l(p)\,.
\end{equation}
Let $\tilde{\delta}(\vec x)$ represent the density fluctuations in real space. The two-point correlation function $F(\vec{\Delta},\vec{X})$ \cite{Rath:2014cka} in real space can be expressed as
\begin{equation}\label{eq:TPC_real_space}
    F(\vec{\Delta},\vec{X})=\langle\tilde{\delta}(\vec{x})\tilde{\delta}(\vec{x}')\rangle\,. 
\end{equation}
where $\Delta = \vec{x} - \vec{x}^{\prime}$ and $\vec{X} = (\vec{x}+\vec{x}^{\prime})/2$. The correlation function in the Fourier space can be expressed as,
\begin{equation}\label{eq:TPC_Fourier_space}
    \langle\delta(\vec{p})\delta^*(\vec{p}')\rangle=\int\frac{d^3X}{(2\pi)^3}\frac{d^3\Delta}{(2\pi)^3}e^{i(\vec{p}+\vec{p}')\cdot\Delta/2}e^{i(\vec{p}-\vec{p}')\cdot\vec{X}}F(\vec{\Delta},\vec{X})\,
\end{equation}
An inhomogeneous model of power spectrum must necessarily depend on $\vec{X}$ in real space. Hence, for such a model, $ F(\vec{\Delta},\vec{X})$ cannot be independent of $\vec{X}$. In such a case, the two-point correlation function can be written, up to the leading order as \cite{Rath:2014cka},
\begin{equation}\label{eq:Inhomogeneous_power_spectrum}
    F(\vec{\Delta},\vec{X})=f_1(\Delta)+\hat{\lambda}\cdot\vec{X}\left(\frac{1}{\eta_0}f_2(\Delta)\right)
\end{equation}
Here, $\hat{\lambda}$ is the unit vector pointing along the preferred direction of anisotropy(unit vector along the $z$-axis in our setup, because we have aligned the modulation along $z$). The scalar functions $f_1(\Delta)$ and $f_2(\Delta)$ are defined as the Fourier transforms of the isotropic power spectrum $P_{iso}(p)$ and the anisotropic spectral function $g(p)$, respectively. The factor $\eta_0$ denotes the conformal time at the present epoch, introduced here to ensure the position term $\vec{X}/\eta_0$ remains dimensionless~\cite{Rath:2014cka}.

To quantify the cumulative effect of these
anisotropic correlations, we have used the direction-dependent statistic $S_H(L)$ defined as the sum of the cross-correlations between adjacent multipoles up to a maximum $\ell$
\begin{equation}\label{eq:statistic}
    S_H(L)=\sum_{\ell=2}^LC_{\ell,\ell+1} .
\end{equation}
where
\begin{equation}\label{eq:Cl_correlation}
    C_{\ell,\ell+1}=\frac{\ell(\ell+1)}{2\ell+1}\sum_{m=-\ell}^{\ell}a_{\ell m}a_{\ell+1,m}^{*} .
\end{equation}
$C_{\ell,\ell+1}$ represents the kind of covariance between harmonic modes $\ell$ and $\ell+1$ sharing the same m (projected along the preferred axis). Using $a_{\ell,-m}=(-1)^m \,a^*_{\ell m}$, it can be shown that $C_{\ell,\ell+1}$ is always real.

This statistic $S_H(L)$ is theoretically motivated by the form of a dipole modulation: a real-space dipole modulation $\Delta T(n)\to \Delta T(n)[1 + A\,(\hat{p}\cdot n)]$ predicts that spherical modes are coupled with $\Delta\ell=1$, with coupling strength proportional to the modulation amplitude A. In essence, $S_H(L)$ accumulates the contributions of all such dipole-induced mode couplings from the quadrupole $(\ell=2)$ up to a scale L, serving as an overall measure of hemispherical anisotropy power. In the absence of any anisotropy, one would expect $S_H$ to be consistent with zero (within cosmic variance). In our inhomogeneous-inflation model, however, we predict a growing $S_H(L)$ at low multipoles, reflecting that large-scale modes carry a coherent dipolar asymmetry.
 
This correlation has been parameterised as \cite{Rath:2014cka},
\begin{equation}\label{eq:alm_correlation}
    \langle a_{\ell m}a_{\ell'm'}^*\rangle=C_\ell\delta_{\ell\ell'}\delta_{mm'}+A(\ell,\ell')
\end{equation}
where
\begin{equation}\label{eq:cl}
    C_\ell=(4\pi)^2\frac{9T_0^2}{100}\int_0^\infty p^2dpj_\ell^2(p\eta_0)P_{iso}(p)\,,
\end{equation}
\begin{equation}\label{eq:All}
A(\ell,\ell')=A_1(\ell,\ell')+A_2(\ell,\ell')\,,
\end{equation}
\begin{figure}[!t]
  \centering
\includegraphics[width=1.0\textwidth]{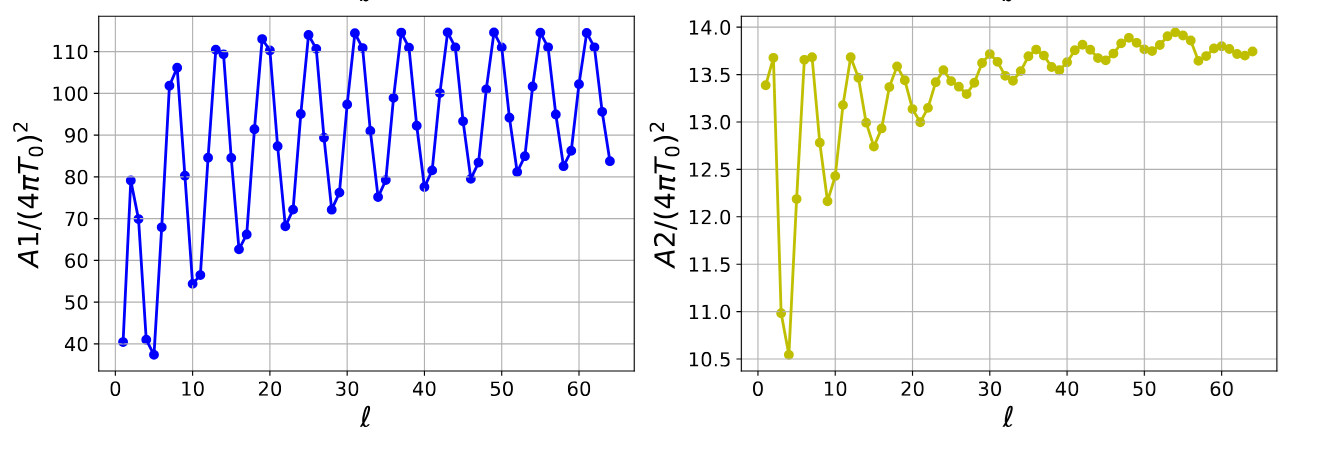}
  \caption{Plot of full expressions of $A_1(\ell,\ell+1)/(4\pi T_0)^2$ and $A_2(\ell,\ell+1)/(4\pi T_0)^2$ as functions of $\ell$, including the prefactors. 
    The oscillatory structure in the plots arises from the angular momentum coupling terms and reflects the detailed multipole dependence of anisotropic contributions in the theoretical model.}
  \label{fig:myimage}
\end{figure}
\begin{equation}\label{eq:A1}
A_{1}(\ell,\ell')=(-i)^{\ell-\ell'+1}2\pi\delta_{\ell',\ell+1}\delta_{m'm}(4\pi T_{0})^{2}N_{\ell m}N_{\ell 'm}I_{\ell}\int dpp g(p)\Theta_{\ell}(p)\Theta_{\ell'}(p)\,,
\end{equation}
$N_{\ell m}=\sqrt{\frac{(2\ell+1) (\ell-m)!}{4\pi (\ell+m)!}}$,  $I_\ell=-2\left[\delta_{0,m}-\frac{(\ell+m)!}{(\ell-m)!}\right]$
for $\ell'=\ell+2n+1, n=0,1,2,....$ and $m>0$ and
\begin{equation}\label{eq:A2}
    \begin{aligned}A_{2}(\ell,\ell')&=\quad-\delta_{\ell',\ell+1}\delta_{m'm}\sqrt{\frac{(\ell-m+1)(\ell+m+1)}{(2\ell+1)(2\ell+3)}}(4\pi T_{0})^{2}\\&\times\quad\int dpp^{2}g(p)\left[\Theta_{\ell'}(p)\frac{d}{dp}\Theta_{\ell}(p)-\Theta_{\ell}(p)\frac{d}{dp}\Theta_{\ell'}(p)\right] .\end{aligned}
\end{equation}
The theoretical link between early-universe inhomogeneity and the observed hemispherical power asymmetry is established by adjacent-$\ell$ mode coupling, where the anisotropic contribution $A(\ell,\ell')$ is non-zero only when $\ell'=\ell\pm1$. 
We restrict our analysis to multipole $\ell<64$ for which a significant signal of hemispherical anisotropy has been observed. Hence we can approximate the transfer function considering only the dominating contribution from the Sachs-Wolfe effect \cite{Gorbunov:2011zzc}. So we can write
\begin{equation}
    \Theta_\ell(p)=\frac{3}{10}j_\ell(p\eta_0)\,.
\end{equation}
While dealing with the transfer function, we have approximated that the conformal time at present $(\eta_0)$ is much greater than the conformal time at recombination $(\eta_d)$.
When $\kappa$ is small, $p=p^{\prime}=p^{\prime\prime}$ which implies $ g\left(p\right)=\frac{\alpha H^{2}\kappa\cos\omega}{p^{3}}[1-\cos\left(\frac{2p}{Ha_{I}}\right)]$. The integrals mentioned in Eqns. \ref{eq:A1} and \ref{eq:A2} have been solved numerically. While evaluating the integrals, we substitute $p\eta_0 = x$, as a result of which the function g(p) becomes $g(x)=\frac{2\eta_0\kappa\alpha\cos\omega}{x}\sin\left(\frac{2x}{\eta_0Ha_{I}}\right)$.

We performed the integrals for a special case (among many possible scenarios) \cite{Rath:2013bfa}, where $q = \frac{2}{\eta_0 H a_I} = 1$. The solution to the horizon problem requires that the conformal time elapsed during inflation should be approximately equal to (or larger than) the time elapsed after inflation. The total conformal time elapsed since the beginning of inflation, denoted by $\eta_0$, should therefore be approximately equal to $2/a_I H$ and hence agrees with our choice,
$2/\eta_0 a_I H = 1$.
We point out that since we are demanding that the effects of the early phase of inflation, during which the universe is inhomogeneous, are observable today, it is reasonable for us to require approximate equality.
We take the limit $|p \eta|<< 1$, because a perturbation with wave number $\vec{p}$ will leave the horizon when $p|\eta|< 1$, thereby obtaining the final values of integrals for different $\ell$ values as shown in Figure \ref{fig:myimage}.

Ideally, we would compare the theoretical predictions in Eqs.~\ref{eq:alm_correlation}--\ref{eq:A2} directly with the observed covariance matrix of the sky. However, observational reality prevents the direct extraction of the true $a_{\ell m}$ coefficients required to evaluate these equations. The application of a Galactic mask $W(\hat{n})$ destroys the orthogonality of the spherical harmonics, causing the observed coefficients $\tilde{a}_{\ell m}$ to be a convolution of the true coefficients and the mask's mode-coupling kernel. Inverting this system to recover the true phases of individual $a_{\ell m}$ modes is mathematically ill-posed without strong priors. To overcome this observational barrier, we adopt a forward-modeling approach. We recognize that our physical model manifests effectively as a dipole modulation of the temperature field. Consequently, we bridge the theoretical derivation to the data analysis by employing the standard dipole modulation estimator. This estimator parameterizes the theoretical $\Delta \ell = 1$ coupling derived above in terms of the power spectrum and the modulation amplitude $A$. Thus, for the observational analysis, we evaluate the covariance using the established relation:
\begin{equation}
\langle a_{\ell m} a^*_{\ell' m'} \rangle = A (C_{\ell+1} + C_{\ell}) \delta_{m' m}
\left[
\sqrt{ \frac{(\ell - m + 1)(\ell + m + 1)}{(2\ell + 1)(2\ell + 3)} } \, \delta_{\ell', \ell+1}
\right]
\label{eq:alm}
\end{equation}
This equation has been derived in \cite{Pranati_K_Rath_2013} using the dipole modulation model decomposing the two-point correlation function of the spherical harmonic coefficients $\langle a_{\ell m} a^*_{\ell' m'} \rangle$ into an isotropic term 
\begin{equation}
\langle a_{\ell m} a^*_{\ell' m'} \rangle_{\mathrm{iso}} = C_{\ell} \delta_{\ell \ell'} \delta_{m m'}
\label{eq:alm_iso}
\end{equation}
and a dipole-induced modulation term
\begin{equation}
\langle a_{\ell m} a^*_{\ell' m'} \rangle_{\mathrm{dm}} = A C_{\ell'} \xi^{0}_{\ell m; \ell' m'} + A C_{\ell} \xi^{0}_{\ell' m'; \ell m}
\label{eq:alm_dm}
\end{equation}
where
\begin{equation}
\begin{aligned}
\xi^{0}_{\ell m; \ell' m'} 
&= \int d\Omega \, Y_{\ell m}^{*}(\hat{n}) \, Y_{\ell' m'}(\hat{n}) \cos \theta \\
&= \delta_{m' m} \left[
\sqrt{ \frac{(\ell - m + 1)(\ell + m + 1)}{(2\ell + 1)(2\ell + 3)} } \, \delta_{\ell', \ell+1}
+ \sqrt{ \frac{(\ell - m)(\ell + m)}{(2\ell + 1)(2\ell - 1)} } \, \delta_{\ell', \ell-1}
\right]
\end{aligned}
\label{eq:xi_integral}
\end{equation}

For concreteness, we fix  A=0.072, consistent with earlier measurements \cite{2014,Hoftuft_2009,refId1}, and postpone a tailored inference using Planck PR4 \texttt{Commander} data to future work. Here, we adopt the well-established idea that perturbations generated during the early inflationary phase may re-enter the horizon during the matter- or radiation-dominated phases \cite{Das_2021}. These modes may have crossed the horizon very early during inflation. This justifies the approximation $p = Ha_I$. As explained earlier, it is expected that anisotropies and inhomogeneities in the Universe will disappear very early during the inflationary phase of expansion \cite{Wald_Robert_1983}.

\begin{figure}[!t]
  \centering
  \includegraphics[width=\textwidth]{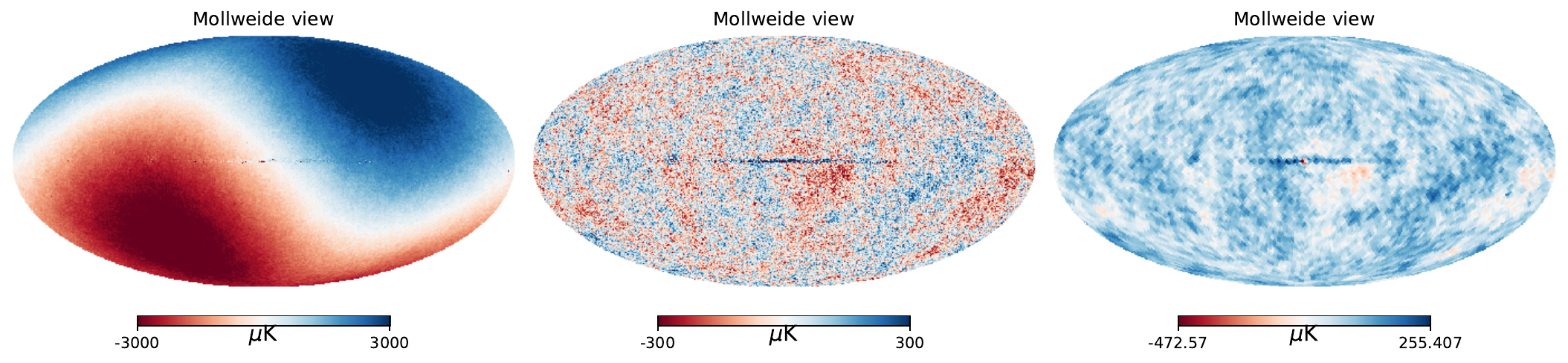}
  \caption{First: Original \texttt{Commander} cleaned PR4 map at $\texttt{nside}=4096$, Second: Monople and dipole removed from the first image, Third: Second image downgraded at $\texttt{nside}=32$}
  \label{fig:cu}

  \vspace{1em}

  \includegraphics[width=\textwidth]{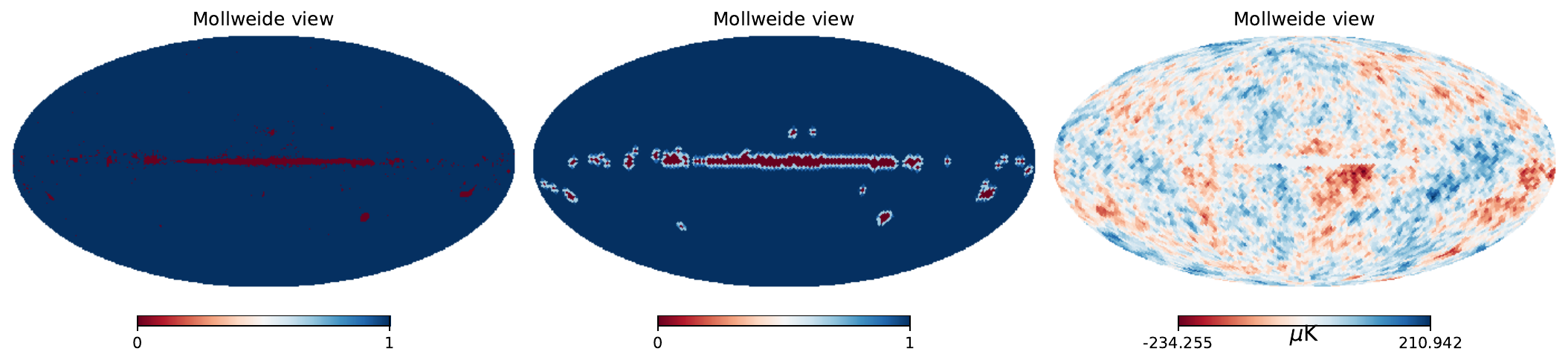}
  \caption{First: Original Common PR3 Inpainting mask at $\texttt{nside}=2048$, Second: Apodized mask at $\texttt{nside}=32$, Third: Masked CMB data at $\texttt{nside}=32$.}
  \label{fig:maskkkkk}
\end{figure}

\section{Data Analysis}
\label{Data Analy}
\subsection{Planck Sky Measurements and Data Processing}
For $C_l$ measurement, the analysis utilizes the PR4 \texttt{Commander} maps, provided at a \texttt{HEALPix} \cite{Gorski_2005} pixel resolution of $\texttt{nside} = 4096$ smoothed with a Gaussian beam of FWHM = 5 arcmin (as shown in first image of Fig \ref{fig:cu}).
We use PR3(2018) \cite{2020} Component Separation Inpainting Common mask in intensity available at $\texttt{nside}=2048$, covering approximately 96.2\% of the sky ($f_{\text{sky}} \approx 96.2\%$)(as shown in first image of Fig \ref{fig:maskkkkk}). We first upgrade this mask to $\texttt{nside}=4096$ by using \texttt{ud\textunderscore grade} from \texttt{HEALPix}, and then smooth it with a Gaussian beam of FWHM = 30 arcmin before applying a cutoff of 0.7. We then downgrade it to $\texttt{nside} = 32$ and apodize it to reduce mode coupling using \texttt{nmt.mask\textunderscore apodization} functionality of \texttt{pymaster}, which is \texttt{python} package of \texttt{NaMaster} \cite{Alonso_2019}. Here, the apodization radius is 3\degree and type of apodization used is C2 (cosine-squared tapering) (as shown in second image of Fig \ref{fig:maskkkkk}).

\begin{figure}[!t]
  \centering
  \begin{minipage}[b]{0.48\textwidth}
    \centering
    \includegraphics[height=6.5cm]{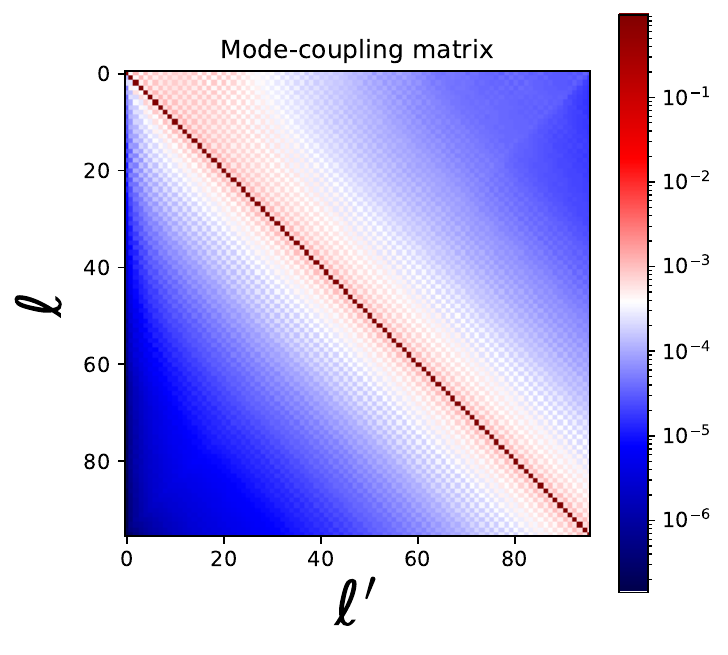}
    \caption{Mode-coupling matrix $M_{\ell \ell'}$ derived from the mask used in the analysis.}
    \label{fig:img2}
  \end{minipage}
  \hfill
  \begin{minipage}[b]{0.5\textwidth}
    \centering
    \includegraphics[height=6cm]{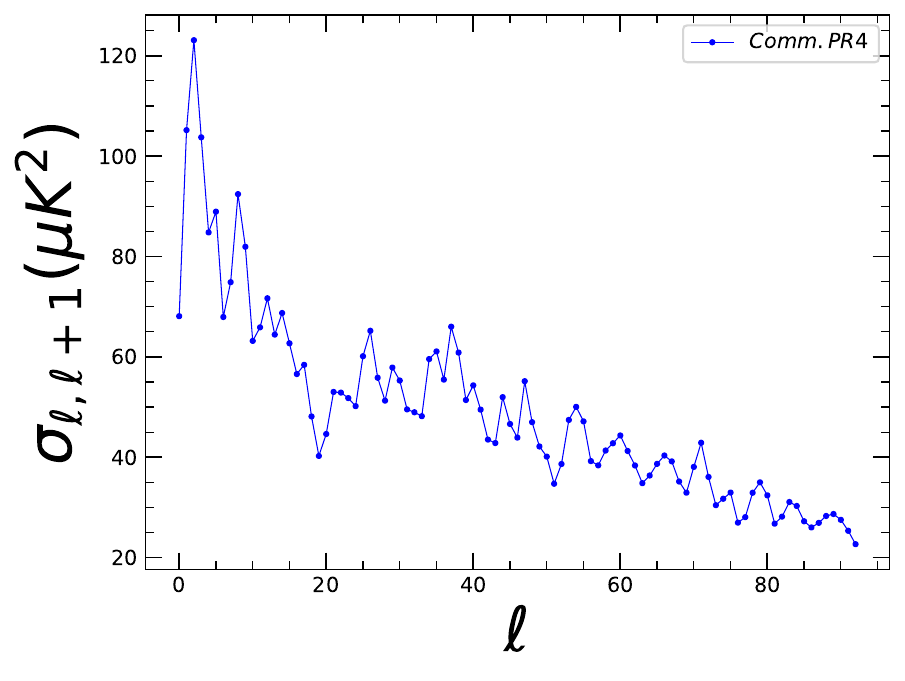}
    \caption{Standard deviation $\sigma_{\ell,\ell+1}$ of $C_{\ell,\ell+1}$ statistic computed from the Planck PR4 \texttt{Commander} map.}
    \label{fig:img1}
  \end{minipage}
  \label{fig:sidebyside}
\end{figure}

We remove the monopole and dipole from CMB map by using \texttt{healpy.remove\textunderscore dipole()} method (as shown in the second image of Fig. \ref{fig:cu}) \citep{refId0}.
For large-scale analysis, the map is downgraded to $\texttt{nside} = 32$ (as shown in the third image of Fig. \ref{fig:cu}) using harmonic space transformations:
\begin{equation}
a_{\ell m}^{\mathrm{out}}
=
\frac{b_{\ell}^{\mathrm{out}}\,p_{\ell}^{\mathrm{out}}}
     {b_{\ell}^{\mathrm{in}}\,p_{\ell}^{\mathrm{in}}}
\,a_{\ell m}^{\mathrm{in}}
\end{equation}
where superscripts ’in’ and ’out’ stand for input (full resolution) and output (low resolution)
maps, respectively and
$b_\ell$ represents the Gaussian beam and $p_\ell$ the HEALPix pixel window functions. We then apply the apodized mask to this downgraded map (as shown in third image of Fig \ref{fig:maskkkkk}) and obtain binned $C_\ell$ using \texttt{NaMaster} (using \texttt{MASTER algorithm}), used to find unbiased angular power spectrum on the sphere  especially in the presence of incomplete sky coverage (i.e. masks) and spin fields. The  mode-coupling matrix (MCM) \cite{Garc_a_Garc_a_2019}, denoted by $M_{\ell \ell'}$ employed in the process is shown in Fig \ref{fig:img2}. It encodes how masking the sky couples different spherical harmonic modes $\ell$ and $\ell'$. It depends entirely on the geometry of the mask and the binning scheme. \texttt{decouple\textunderscore cell} fuctionality of \texttt{NaMaster} inverts this matrix (along with noise corrections etc.) to recover the unbiased power spectrum.
\begin{equation}
\hat{C}_\ell = \sum_{\ell'} M^{-1}_{\ell \ell'} \, \tilde{C}_{\ell'}
\end{equation}
We feed this $\hat{C}_\ell$ in Eqn. \ref{eq:alm} to obtain $C_{\ell,\ell+1}$ and hence $S_H(L)$. 


\subsection{\texorpdfstring{$\chi^2$}{Chi-square}-minimization}
We next extract the values of the model parameters, $\alpha$, $\kappa$ and $\omega$ for which the theoretical value of $S_{H}(L)$ matches the data value in the multipole range 2-63 using the $\chi^2$-minimization procedure. Here $\chi^2$ is defined as
\begin{equation}
\chi^2 = \sum_L \frac{\left[ S_H^{\text{theory}}(L) - S_H^{\text{data}}(L) \right]^2}{\left[ \delta S_H^{\text{data}}(L) \right]^2}
\label{S_H}
\end{equation}
where $S_H^{\text{theory}}(L)$ is the theoretical estimate of the statistic in a said multipole range using the inhomogeneous model, $S_H^{\text{data}}(L)$ the bias corrected estimate and $\delta S_H^{\text{data}}(L)$ the corresponding error. In this formulation, the summation is evaluated over the specific, discrete five multipole bins that are explicitly defined in Table \ref{tab:com_statistics} (excluding $\ell = 2$ to $\ell = 63$). To facilitate the estimation of the data values for \(S_H(L)\), the direction parameters were held constant across all included bins. In particular, we adopted the fixed Galactic coordinates \((l, b) = (221^\circ, -21^\circ) \pm 31^\circ\) \cite{refId1} as the reference direction for the entirety of the binning scheme. To estimate $\delta S_H^{\text{data}}(L)$, we use a measure of cosmic variance, $\sigma_{\ell, \ell+1}$, given by, 
\begin{equation}
\sigma_{\ell, \ell+1} = \sqrt{\mathrm{Var}(C_{\ell,\ell+1})} \approx \sqrt{\langle C_{\ell,\ell+1}^2 \rangle - \langle C_{\ell,\ell+1} \rangle^2 }\approx \frac{\ell (\ell+1)}{\sqrt{2\ell+1}}\, \sqrt{C_\ell C_{\ell+1}}
\end{equation}
The resulting plot for the data set used in this paper is shown in Fig \ref{fig:img1}. 
Since the contribution from anisotropic term was very small, it has been neglected in the expression of $\sigma_{\ell, \ell+1}$ for plotting purposes.

\begin{figure}[!t]
  \centering
  \includegraphics[width=\textwidth]{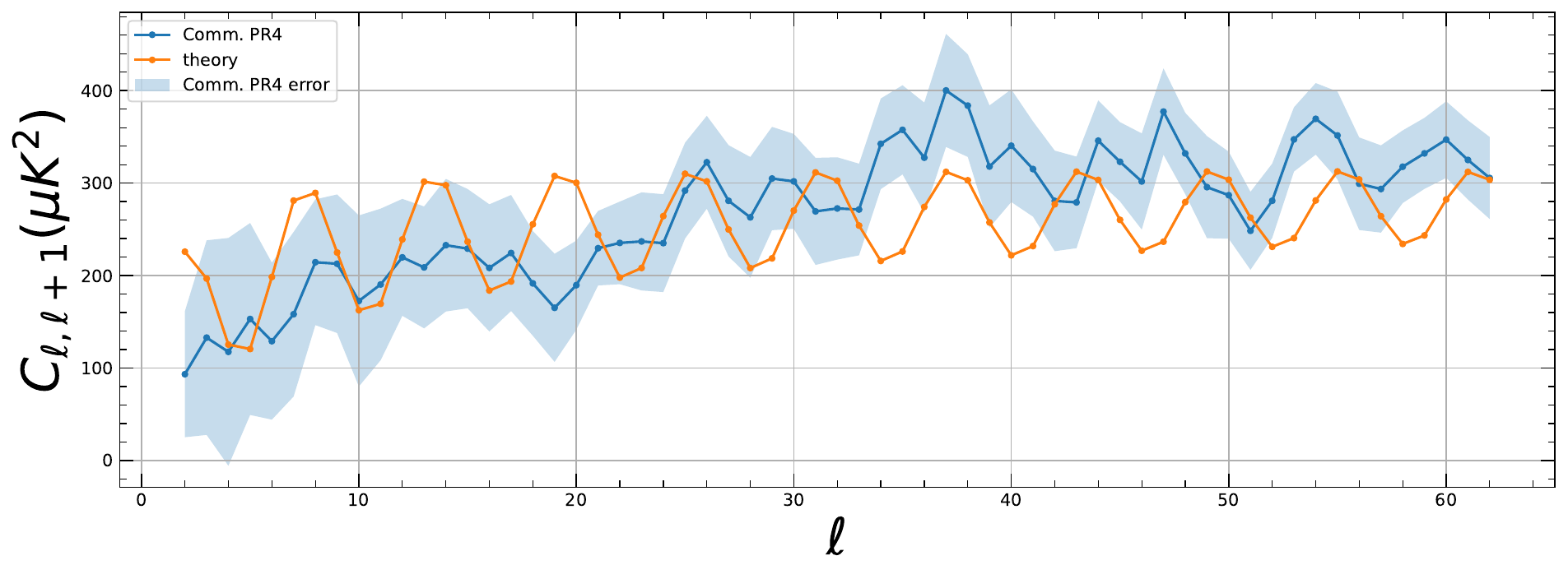}
  \caption{Comparison of $C_{\ell,\ell+1}$ between data and theory for best fit parameters $\alpha, \kappa$ and $\omega$ obtained from $\chi^2$ minimisation of $S_H(L)$ in the multipole range 2--63.}
  \label{fig:huhgygft}
\end{figure}

\section{Results and Discussions}
The model’s specific sinusoidal perturbation produces a characteristic $\ell$ dependence: Figure~\ref{fig:huhgygft} which compares the estimated correlation $C_{\ell,\ell+1}$ from the Planck PR4 \texttt{Commander} map (blue solid line) for different $\ell$-values with the theoretical prediction for the best–fit parameters
$\alpha$, $\kappa$ and $\omega$ (orange line) in the multipole range 2--63. The blue shaded band shows the corresponding $1\sigma$ uncertainty on the Planck PR4 \texttt{Commander} data.
To further investigate the multipole dependence of the signal, we present the binned values of the statistic $S_H$ in Table \ref{tab:com_statistics}. This provides a more granular view of the contributions to the cumulative statistic. The analysis indicates that the dominant contribution to the overall signal arises from the lower multipole bins (e.g., $\ell=2-22$), which is consistent with the large-scale nature of the hemispherical asymmetry anomaly. We restrict our analysis to multipoles $\ell < 64$, where hemispherical asymmetry signals are strongest in observations.

\begin{figure}[ht]
  \centering
  \includegraphics[width=0.78\textwidth]{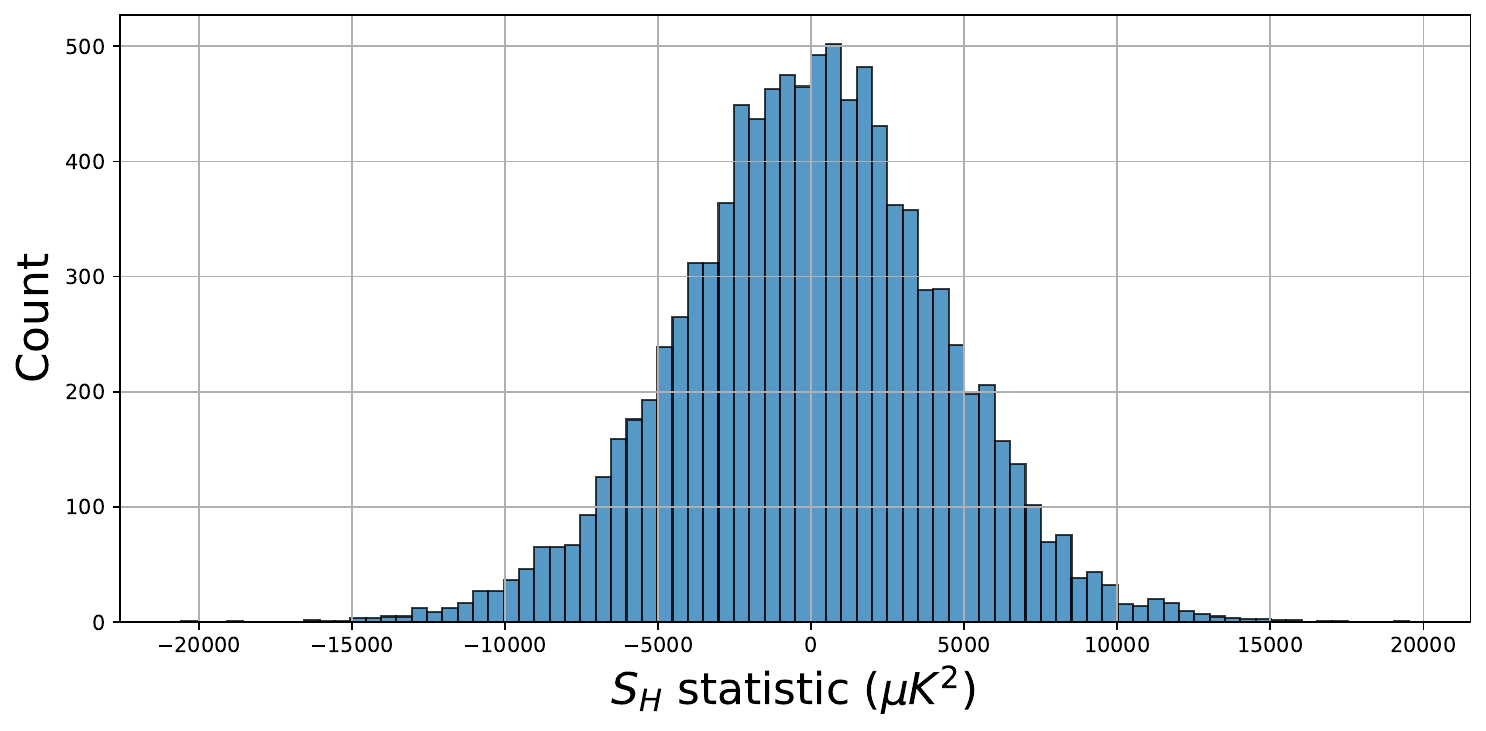}
  \caption{Histogram of the $S_H$ statistic from 10000 ideal CMB realisations for the multipole range 2--63.}
  \label{fig:saitc}
\end{figure}

In order to determine the significance of the hemispherical anisotropy in data, 
we generated an ensemble of 10{,}000 ideal CMB temperature maps using the best-fit cosmological parameters \cite{Tristram_2024} from the Planck PR4 \texttt{Commander} likelihood. These parameters include $H_0 = \mathbf{67.64\pm0.52}~\mathrm{km\,s^{-1}\,Mpc^{-1}}$, $\Omega_b h^2 = \mathbf{0.02224\pm0.00025}$, $\Omega_c h^2 = \mathbf{0.1183\pm0.0024}$, scalar amplitude $A_s = \mathbf{(2.16\pm0.13)\times 10^{-9}}$, scalar spectral index $n_s = \mathbf{0.9678\pm0.0072}$, and optical depth $\tau = \mathbf{0.0753\pm0.0322}$. We used the \texttt{CAMB} code \cite{Lewis_2000, Howlett_2012} to compute the theoretical angular power spectrum $C_\ell$ up to $\ell_{\max} = 95$, ensuring consistency with large-scale anisotropy analysis. For each simulated map, we computed the corresponding $S_H$ value, and compiled the results into a distribution. The resulting histogram, shown in Fig.~\ref{fig:saitc}, depicts the distribution of $S_H$ values over 10000 valid maps (maps that existed in the input directory). The distribution is approximately Gaussian, centered near the mean theoretical expectation for the isotropic $\Lambda$CDM model, and provides a baseline for estimating the $p$-value when comparing real CMB data against simulations. The corresponding p-value for the range, $\ell = 2-63$, is given in Table \ref{tab:com_statistics}, along with the p-values of other bins.

\begin{table}[!t]
\centering
\begin{tabular}{ccccccc}
\hline
\textbf{Range} & $02-22$ & $23-33$ & $34-44$ & $45-55$ & $56-63$ & $2-63$ \\
\textbf{Statistic} & $6650 \pm 2200$ & $3498 \pm 2130$ & $1624 \pm 1280$ & $849 \pm 735$ & $410 \pm 405$ & $13041 \pm 4210$ \\
\textbf{p-value} & $0.003$ & $0.10$ & $0.20$ & $0.25$ & $0.32$ & $0.002$ \\
\hline
\end{tabular}
\caption{The values of statistic $S_H$(L) and their $1\sigma$ uncertainties in 6 bins from PR4 \texttt{Commander} with fixed direction parameters at $(l,b) = (221^{\circ}, -21^{\circ}) \pm 31 \degree$.}
\label{tab:com_statistics}
\end{table}

In the theoretical expression for the \( C_{l,l+1} \) correlation, the perturbative contribution enters only through the combined parameter $\alpha \kappa \cos \omega$, so the data constrains this product rather than the individual parameters separately. For simplicity, and without loss of generality for the present fit, we set the phase angle to \(\omega = 0\), reducing the relevant amplitude to the single combination $\alpha \kappa$. To determine the amplitude of the inhomogeneous perturbation, we initially perform a \(\chi^2-\) fit (Eqn. \ref{S_H}) comparing the theoretical template \(S_H^{\text{theory}}(L)\) to the observed statistic \(S_H^{\text{data}}(L)\) without applying any bias corrections. The \(\chi^2\) is evaluated by summing over the individual multipole bins listed in Table \ref{tab:com_statistics}, thereby incorporating the binned scale dependence of the signal into the fit. This uncorrected fit yields a raw amplitude of \((\alpha \kappa)_{\text{raw}} = (0.89 \pm 0.27) \times 10^{-7}\text{Mpc}^{-1}\).



The estimator $S_H$ is known to acquire a small but non-negligible bias in the presence of masking and related effects, an issue that has been examined in detail in earlier simulation-based studies such as \cite{Rath:2014cka}. For the present analysis, we reassessed this effect using Planck PR4 \texttt{Commander}-based simulations processed through the same analysis pipeline as the data. Specifically, we generated an ensemble of ideal CMB temperature realizations using the best-fit PR4 cosmological parameters, convolved each realization with the appropriate beam, applied the same downgraded and apodized mask, and then processed the maps through the same NaMaster pipeline used for the data analysis. For each simulated realization, we computed the statistic $S_H$ over the multipole range $2\leq \ell \leq 63$ after fixing the same directional parameters as in the data. We then fitted the theoretical template to the resulting simulated values in order to obtain the sampling distribution of the inferred parameter combination $\alpha\kappa$. This exercise shows that the estimator exhibits only a mild residual bias. We find that, over the multipole interval $2\leq \ell \leq 63$, the resulting fractional bias in $S_H$ is approximately $8\%$, matching with the previous analysis \cite{Rath:2014cka, Kothari:2015tqa, Ghosh:2015qta}. Accordingly, we apply an explicit bias correction to the measured $S_H$ in each bin prior to parameter inference. The bias-corrected value of the quantity $\alpha\kappa$, inferred from the corrected $S_H$ statistic is $(0.97\pm0.31)\times10^{-7}\text{Mpc}^{-1}$. To establish a representative baseline for our model, we take the bias corrected reference values to be $\alpha = (0.44 \pm 0.13) \times 10^{-2}$, $\kappa = (2.19 \pm 0.67) \times 10^{-5} \, \text{Mpc}^{-1}$. Here we have assumed that the percentage error in these two parameters is the same. 

Our background metric has a small inhomogeneity, which is being treated as a perturbation. Since $\kappa H_0^{-1}<<1$, this acts as a superhorizon adiabatic perturbation and can affect the CMB quadrupole and octopole. In order to be consistent with observations, we impose the following constraints \cite{Erickcek_2008, Tiwari_2022} from measurements of the CMB quadrupole and octupole
\begin{equation}
\bigl|\alpha_{\mathrm{dec}}\sin\omega\bigr|
   \;\le\;
   \dfrac{5.8\,Q}{\bigl(\kappa \chi_{\mathrm{dec}}\bigr)^{2}}
   \label{c1}
\end{equation}
\begin{equation}
\bigl|\alpha_{\mathrm{dec}}\cos\omega\bigr|
   \;\le\;
   \dfrac{32\,\mathcal{O}}{\bigl(\kappa \chi_{\mathrm{dec}}\bigr)^{3}}
   \label{c2}
\end{equation}
Here, the subscript "dec" indicates quantities evaluated at the time of decoupling. $\chi_{dec}$ is the comoving distance to decoupling. Here, $\alpha_{dec}=0.937\alpha$(comes from $\Psi_{dec}=0.937\Psi$ \cite{Erickcek_2008}). We use the latest Planck 2020 \cite{refId0} values $Q = 3\sqrt{C_2} \lesssim 1.7\times10^{-5}$ and ${\cal O} = 3\sqrt{C_3} \lesssim 3.1 \times 10^{-5}$, three times the measured rms values of the quadrupole and octupole, as $3\sigma$ upper limits. These bounds stem from the Grishchuk-Zel'dovich effect, which dictates how super-horizon perturbations generate large-scale temperature anisotropies. We see that our perturbation easily satisfies both the constraints with the parameter values above obtained. This consistency is crucial, as it confirms that the proposed perturbation is physically viable and does not conflict with the well-constrained, large-scale structure of the CMB.

\begin{figure}[!t]
  \centering
  \includegraphics[width=0.68\textwidth]{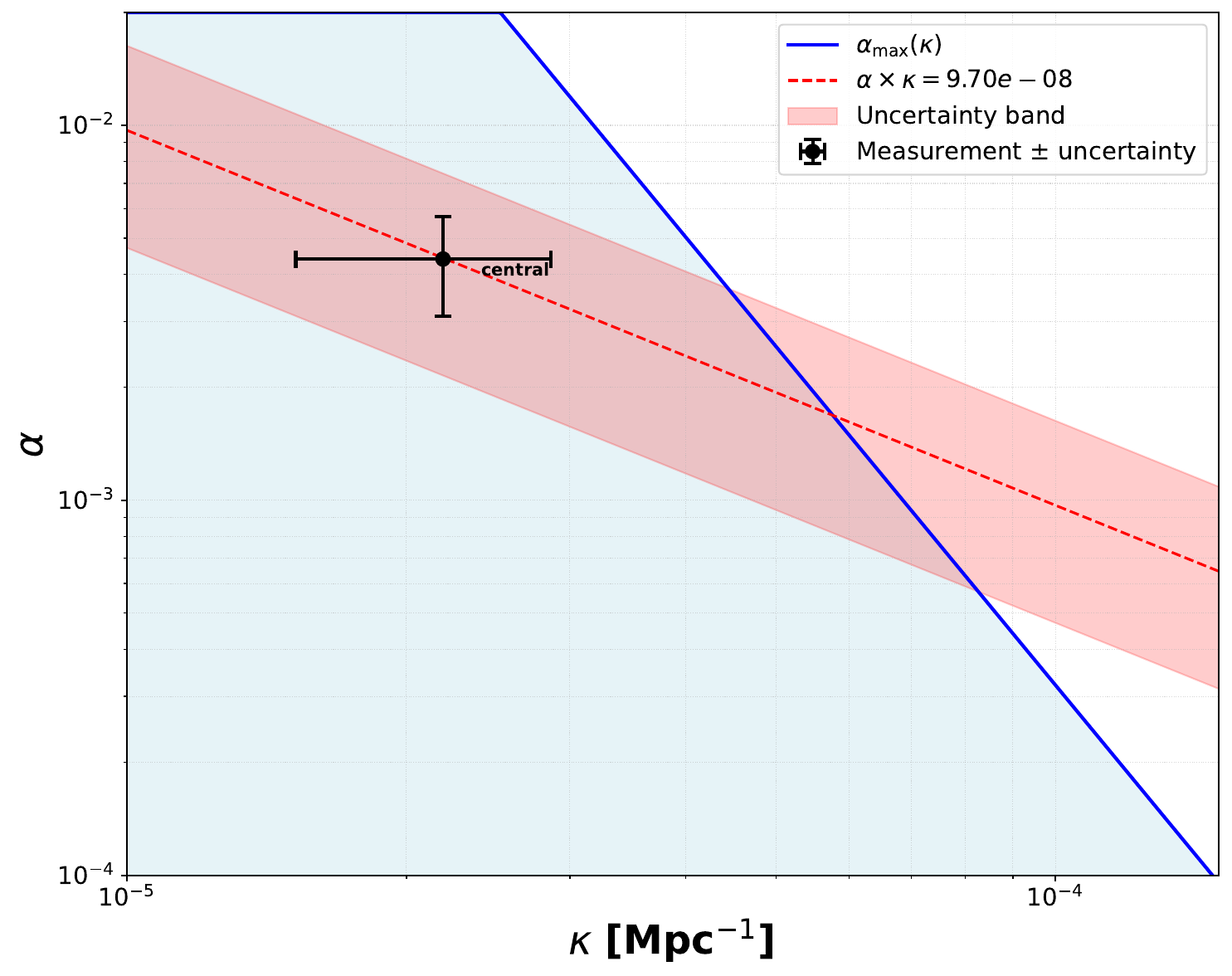}
  \caption{Constraints on the anisotropy parameters $\alpha$ and $\kappa$. The blue solid curve shows the upper limit $\alpha_{\text{max}}(\kappa)$ derived from the quadrupole and octupole constraints, with the light blue region indicating the allowed parameter space. The red dashed hyperbola shows the best-fit constraint $\alpha \kappa \approx 0.97 \times 10^{-7}\text{Mpc}^{-1}$ derived from the hemispherical asymmetry $\chi^2$ analysis, with the red band representing the $1\sigma$ uncertainty. The black point indicates the representative central measurements used in this work.}
  \label{fig:bound}
\end{figure}

Figure~\ref{fig:bound} summarizes the constraints on the anisotropy parameters. The blue solid curve represents the theoretical exclusion limit $\alpha_{\text{max}}(\kappa)$, derived by translating the observed Planck quadrupole ($Q$) and octupole ($\mathcal{O}$) amplitudes into upper bounds on super-horizon perturbations using Eqs.~\ref{c1} and \ref{c2}. The light-blue shaded region below this curve defines the parameter space consistent with these low-$\ell$ constraints. The red dashed curve represents the constraint derived from our hemispherical asymmetry analysis. Since the dipole modulation signal in our model scales with the product $\alpha \kappa$ (for the phase $\omega=0$), the data primarily constrain this combination. The curve shows the best-fit hyperbola $\alpha \kappa \approx 0.97 \times 10^{-7}\text{Mpc}^{-1}$, obtained via $\chi^2$ minimization of the mode-coupling statistic $S_H(L)$ over the multipole range $\ell = 2-63$. The red shaded band indicates the corresponding $1\sigma$ uncertainty region propagated from this fit. Finally, the black point marks the central values for the specific representative case used in our analysis, with error bars illustrating the relative uncertainty implied by the fit.

An in-built assumption 
in our formalism is that the inhomogeneity will quickly decay during the initial stages 
of inflation. Such a decay is expected and has been explicitly demonstrated for a class of anisotropic models ~\cite{Wald_Robert_1983}. This is likely to 
explain the scale dependence of the HPA and may also satisfy Gaussianity 
constraints, as observed in Byrnes et al. \cite{Byrnes_2016, Byrnes_2015}. However, so far we have not really 
determined the non-Gaussianity contribution within our model. This needs more work, which is postponed to future research. 

\section{Conclusion}
Our findings suggest that a brief inhomogeneous phase of inflation is a possible mechanism for the observed hemispherical power asymmetry of the CMB. By assuming that the metric at an early stage of inflation shows a small deviation from homogeneity, implemented with a long-wavelength perturbation, we obtain a direction-dependent power spectrum that naturally produces hemispheric differences in CMB power. The resulting signature, the nonzero correlations between adjacent multipoles $(\ell, \ell+1)$, is found in the Planck data with a statistical significance of about 3 sigmas. The observed scale-dependence of these correlations also agrees with theoretical expectations. Hence, the early inhomogeneous phase of inflation opens up a promising avenue to explain small deviations from isotropy seen in cosmological data. 

We acknowledge, however, that constructing concrete inflationary realizations of this phenomenology is theoretically challenging. As shown in recent literature (e.g., Byrnes et al. 2016 \cite{Byrnes_2016}), models relying on super-horizon modulation often face tension between generating sufficient asymmetry amplitude and satisfying Planck's tight constraints on non-Gaussianity ($f_{NL}$). Resolving this typically requires significant fine-tuning or scale-dependent bispectra to suppress non-Gaussianity on observable scales.
Despite these challenges, the inhomogeneous 
framework presented here offers distinct, testable predictions and follows naturally from the Big Bang paradigm.


In summary, we have used the in-in formalism \cite{Weinberg:2005vy} to analyse an early phase of inflation  during which the Universe may show some departure from isotropy and homogeneity. We have shown that a small inhomogeneity, which can be treated perturbatively, can explain the observed hemispherical anisotropy \cite{2004ApJ...605...14E}. The basic idea that an early phase of inflation can generate the observed signals of anisotropy has already been explored in the literature. Furthermore, it has also been used to study an anisotropic early phase of inflation \cite{PhysRevD.75.083502, Rath:2013bfa}. However, an inhomogeneous early phase has been treated for the first time in our paper. This is important since such a model is needed to describe the CMB hemispherical anisotropy. Our work provides a concrete implementation of this idea and demonstrates its consistency with
observations. However, considerable more effort is needed in order to build a complete model that incorporates the full matter content during this early phase. Future investigations may further test this mechanism, for instance, by examining
polarization maps or higher-resolution data. If successful, this offers an exciting possibility of testing the very early phase of cosmic evolution through large distance scale effects.

\acknowledgments
We are grateful to an anonymous referee for thorough evaluation and for pointing out technical inaccuracies. We acknowledge the use of Python packages \texttt{matplotlib} \cite{article}, \texttt{scipy} \cite{Virtanen_2020},  \texttt{numpy} \cite{harris2020array}, \texttt{astropy} \cite{2022ApJ...935..167A} and \texttt{healpy} \cite{Zonca2019}
for this analysis. This work used data from the Planck Legacy Archive (PLA), operated by ESA and hosted by the European Space Astronomy Center (ESAC) \footnote{\url{https://pla.esac.esa.int/}}. We also acknowledge the computational facility provided by the Department of Space, Planetary \& Astronomical Sciences \& Engineering (SPASE), Indian Institute of Technology Kanpur.






\bibliographystyle{JHEP}
\bibliography{Ref}

\end{document}

%% file: def.tex
\usepackage[ngerman,english]{babel}
\newif\ifAMStwofonts
\AMStwofontstrue

\def\gsim{~\rlap{$>$}{\lower 1.0ex\hbox{$\sim$}}}

\def\simpropto{\lower.2ex\hbox{$\; \buildrel \propto \over \sim \;$}}
\def\ltsim{\lower.5ex\hbox{$\; \buildrel < \over \sim \;$}}
\def\gtsim{\lower.5ex\hbox{$\; \buildrel > \over \sim \;$}}
\def\ltsim{\lower.5ex\hbox{$\; \buildrel < \over \sim \;$}}
\def\gtsim{\lower.5ex\hbox{$\; \buildrel > \over \sim \;$}}

\def\pmb#1{\setbox0=\hbox{#1}%
\kern-.025em\copy0\kern-\wd0
\kern.05em\copy0\kern-\wd0
\kern-.025em\raise.0433em\box0}

\def\simlt{\lower.5ex\hbox{$\; \buildrel < \over \sim \;$}}
\def\simgt{\lower.5ex\hbox{$\; \buildrel > \over \sim \;$}}

\newcommand{\beq}{\begin{equation}}

\newcommand{\eeq}{\end{equation}}
\def\beqa{\begin{eqnarray}}
\def\eeqa{\end{eqnarray}}
\def\fixit#1{}


%% file: Ref.bib
@article{PhysRevD.42.313,
  title = {Isocurvature baryon perturbations and inflation},
  author = {Mollerach, Silvia},
  journal = {Phys. Rev. D},
  volume = {42},
  issue = {2},
  pages = {313--325},
  numpages = {0},
  year = {1990},
  month = {Jul},
  publisher = {American Physical Society},
  doi = {10.1103/PhysRevD.42.313},
  url = {https://link.aps.org/doi/10.1103/PhysRevD.42.313}
}

@article{Blake_2002,
        doi = {10.1038/416150a},
        url = {https://doi.org/10.1038%2F416150a},
        year = 2002,
        month = {mar},
        publisher = {Springer Science and Business Media {LLC}
        },
        volume = {416},
        number = {6877},
        pages = {150--152},
        author = {Chris Blake and Jasper Wall},
        title = {A velocity dipole in the distribution of radio galaxies},
        journal = {Nature}
}

@article{Secrest_2021,
doi = {10.3847/2041-8213/abdd40},
url = {https://dx.doi.org/10.3847/2041-8213/abdd40},
year = {2021},
month = {feb},
publisher = {The American Astronomical Society},
volume = {908},
number = {2},
pages = {L51},
author = {Nathan J. Secrest and Sebastian von Hausegger and Mohamed Rameez and Roya Mohayaee and Subir Sarkar and Jacques Colin},
title = {A Test of the Cosmological Principle with Quasars},
journal = {The Astrophysical Journal Letters}
}

@article{Zibin:2015ccn,
    author = "Zibin, J. P. and Contreras, D.",
    title = "{Testing physical models for dipolar asymmetry: from temperature to k space to lensing}",
    eprint = "1512.02618",
    archivePrefix = "arXiv",
    primaryClass = "astro-ph.CO",
    doi = "10.1103/PhysRevD.95.063011",
    journal = "Phys. Rev. D",
    volume = "95",
    number = "6",
    pages = "063011",
    year = "2017"
}

@article{Planck:2015igc,
    author = "Ade, P. A. R. and others",
    collaboration = "Planck",
    title = "{Planck 2015 results. XVI. Isotropy and statistics of the CMB}",
    eprint = "1506.07135",
    archivePrefix = "arXiv",
    primaryClass = "astro-ph.CO",
    doi = "10.1051/0004-6361/201526681",
    journal = "Astron. Astrophys.",
    volume = "594",
    pages = "A16",
    year = "2016"
}

@article{Akrami:2014eta,
    author = "Akrami, Y. and Fantaye, Y. and Shafieloo, A. and Eriksen, H. K. and Hansen, F. K. and Banday, A. J. and G\'orski, K. M.",
    title = "{Power asymmetry in WMAP and Planck temperature sky maps as measured by a local variance estimator}",
    eprint = "1402.0870",
    archivePrefix = "arXiv",
    primaryClass = "astro-ph.CO",
    doi = "10.1088/2041-8205/784/2/L42",
    journal = "Astrophys. J. Lett.",
    volume = "784",
    pages = "L42",
    year = "2014"
}

@article{Kothari:2015tqa,
    author = "Kothari, Rahul and Ghosh, Shamik and Rath, Pranati K. and Kashyap, Gopal and Jain, Pankaj",
    url = {https://dx.doi.org/10.3847/2041-8213/abdd40}, 
    title = "{Imprint of Inhomogeneous and Anisotropic Primordial Power Spectrum on CMB Polarization}",
    eprint = "1503.08997",
    archivePrefix = "arXiv",
    primaryClass = "astro-ph.CO",
    doi = "10.1093/mnras/stw1039",
    journal = "Mon. Not. Roy. Astron. Soc.",
    volume = "460",
    number = "2",
    pages = "1577--1587",
    year = "2016"
}

@article{Koivisto:2010fk,
    author = "Koivisto, Tomi S. and Mota, David F.",
    title = "{CMB statistics in noncommutative inflation}",
    eprint = "1011.2126",
    archivePrefix = "arXiv",
    primaryClass = "astro-ph.CO",
    reportNumber = "ITP-UU-10-41, SPIN-10-34",
    doi = "10.1007/JHEP02(2011)061",
    journal = "JHEP",
    volume = "02",
    pages = "061",
    year = "2011"
}

@article{Bennett:2010jb,
    author = "Bennett, C. L. and others",
    title = "{Seven-Year Wilkinson Microwave Anisotropy Probe (WMAP) Observations: Are There Cosmic Microwave Background Anomalies?}",
    eprint = "1001.4758",
    archivePrefix = "arXiv",
    primaryClass = "astro-ph.CO",
    doi = "10.1088/0067-0049/192/2/17",
    journal = "Astrophys. J. Suppl.",
    volume = "192",
    pages = "17",
    year = "2011"
}

@article{Ghosh:2015qta,
    author = "Ghosh, Shamik and Kothari, Rahul and Jain, Pankaj and Rath, Pranati K.",
    title = "{Dipole Modulation of Cosmic Microwave Background Temperature and Polarization}",
    eprint = "1507.04078",
    archivePrefix = "arXiv",
    primaryClass = "astro-ph.CO",
    doi = "10.1088/1475-7516/2016/01/046",
    journal = "JCAP",
    volume = "01",
    pages = "046",
    year = "2016"
}

@article{Adhikari:2014mua,
    author = "Adhikari, Saroj",
    title = "{Local variance asymmetries in Planck temperature anisotropy maps}",
    eprint = "1408.5396",
    archivePrefix = "arXiv",
    primaryClass = "astro-ph.CO",
    reportNumber = "IGC-14-8-2",
    doi = "10.1093/mnras/stu2408",
    journal = "Mon. Not. Roy. Astron. Soc.",
    volume = "446",
    number = "4",
    pages = "4232--4238",
    year = "2015"
}

@article{Axelsson:2013mva,
    author = "Axelsson, M. and Fantaye, Y. and Hansen, F. K. and Banday, A. J. and Eriksen, H. K. and Gorski, K. M.",
    title = "{Directional dependence of $\Lambda$CDM cosmological parameters}",
    eprint = "1303.5371",
    archivePrefix = "arXiv",
    primaryClass = "astro-ph.CO",
    doi = "10.1088/2041-8205/773/1/L3",
    journal = "Astrophys. J. Lett.",
    volume = "773",
    pages = "L3",
    year = "2013"
}

@article{PhysRevD.71.083508,
  title = {Constraints on mode couplings and modulation of the CMB with WMAP data},
  author = {Prunet, Simon and Uzan, Jean-Philippe and Bernardeau, Francis and Brunier, Tristan},
  journal = {Phys. Rev. D},
  volume = {71},
  issue = {8},
  pages = {083508},
  numpages = {12},
  year = {2005},
  month = {Apr},
  publisher = {American Physical Society},
  doi = {10.1103/PhysRevD.71.083508},
  url = {https://link.aps.org/doi/10.1103/PhysRevD.71.083508}
}

@article{Gordon_2007,
   title={Broken Isotropy from a Linear Modulation of the Primordial Perturbations},
   volume={656},
   ISSN={1538-4357},
   url={http://dx.doi.org/10.1086/510511},
   DOI={10.1086/510511},
   number={2},
   journal={The Astrophysical Journal},
   publisher={American Astronomical Society},
   author={Gordon, Christopher},
   year={2007},
   month={feb}, 
   pages={636–640} 
}

@article{Gordon_2005,
   title={Spontaneous isotropy breaking: A mechanism for CMB multipole alignments},
   volume={72},
   ISSN={1550-2368},
   url={http://dx.doi.org/10.1103/PhysRevD.72.103002},
   DOI={10.1103/physrevd.72.103002},
   number={10},
   journal={Physical Review D},
   publisher={American Physical Society (APS)},
   author={Gordon, Christopher and Hu, Wayne and Huterer, Dragan and Crawford, Tom},
   year={2005},
   month={nov}
}

@article{Erickcek_2008,
   title={Superhorizon perturbations and the cosmic microwave background},
   volume={78},
   ISSN={1550-2368},
   url={http://dx.doi.org/10.1103/PhysRevD.78.083012},
   DOI={10.1103/physrevd.78.083012},
   number={8},
   journal={Physical Review D},
   publisher={American Physical Society (APS)},
   author={Erickcek, Adrienne L. and Carroll, Sean M. and Kamionkowski, Marc},
   pages = "083012",
   year={2008},
   month={oct}
}

@article{Weinberg:2005vy,
    author = "Weinberg, Steven",
    title = "{Quantum contributions to cosmological correlations}",
    eprint = "hep-th/0506236",
    archivePrefix = "arXiv",
    reportNumber = "UTTG-01-05",
    doi = "10.1103/PhysRevD.72.043514",
    journal = "Phys. Rev. D",
    volume = "72",
    pages = "043514",
    year = "2005"
}

@article{Rath:2013bfa,
    author = "Rath, Pranati K. and Mudholkar, Tanmay and Jain, Pankaj and Aluri, Pavan K. and Panda, Sukanta",
    title = "{Direction dependence of the power spectrum and its effect on the Cosmic Microwave Background Radiation}",
    eprint = "1302.2706",
    archivePrefix = "arXiv",
    primaryClass = "astro-ph.CO",
    doi = "10.1088/1475-7516/2013/04/007",
    journal = "JCAP",
    volume = "04",
    pages = "007",
    year = "2013"
}

@article{Tiwari_2022,
   title={Superhorizon Perturbations: A Possible Explanation of the Hubble–Lemaître Tension and the Large-scale Anisotropy of the Universe},
   volume={924},
   ISSN={2041-8213},
   url={http://dx.doi.org/10.3847/2041-8213/ac447a},
   DOI={10.3847/2041-8213/ac447a},
   number={2},
   journal={The Astrophysical Journal Letters},
   publisher={American Astronomical Society},
   author={Tiwari, Prabhakar and Kothari, Rahul and Jain, Pankaj},
   year={2022},
   month={jan}, 
   pages={L36} 
}

@article{Jain_2015,
   title={Noncommutative geometry and the primordial dipolar imaginary power spectrum},
   volume={75},
   ISSN={1434-6052},
   url={http://dx.doi.org/10.1140/epjc/s10052-015-3333-9},
   DOI={10.1140/epjc/s10052-015-3333-9},
   number={3},
   journal={The European Physical Journal C},
   publisher={Springer Science and Business Media LLC},
   author={Jain, Pankaj and Rath, Pranati K.},
   year={2015},
   month={mar} 
}

@article{Land_2005,
   title={Examination of Evidence for a Preferred Axis in the Cosmic Radiation Anisotropy},
   volume={95},
   ISSN={1079-7114},
   url={http://dx.doi.org/10.1103/PhysRevLett.95.071301},
   DOI={10.1103/physrevlett.95.071301},
   number={7},
   journal={Physical Review Letters},
   publisher={American Physical Society (APS)},
   author={Land, Kate and Magueijo, João},
   year={2005},
   month={aug} 
}

@ARTICLE{2004PhRvD..69f3516D,
       author = {{de Oliveira-Costa}, Ang{\'e}lica and {Tegmark}, Max and {Zaldarriaga}, Matias and {Hamilton}, Andrew},
        title = "{Significance of the largest scale CMB fluctuations in WMAP}",
      journal = {\prd},
         year = {2004},
        month = {mar},
       volume = {69},
       number = {6},
          eid = {063516},
        pages = {063516},
          doi = {10.1103/PhysRevD.69.063516},
archivePrefix = {arXiv},
       eprint = {astro-ph/0307282},
 primaryClass = {astro-ph},
       adsurl = {https://ui.adsabs.harvard.edu/abs/2004PhRvD..69f3516D}
}

@ARTICLE{2011ApJ...742L..23S,
       author = {{Singal}, Ashok K.},
        title = "{Large Peculiar Motion of the Solar System from the Dipole Anisotropy in Sky Brightness due to Distant Radio Sources}",
      journal = {\apjl},
         year = {2011},
        month = {dec},
       volume = {742},
       number = {2},
          eid = {L23},
        pages = {L23},
          doi = {10.1088/2041-8205/742/2/L23},
archivePrefix = {arXiv},
       eprint = {1110.6260},
 primaryClass = {astro-ph.CO},
       adsurl = {https://ui.adsabs.harvard.edu/abs/2011ApJ...742L..23S}
}

@ARTICLE{2012MNRAS.427.1994G,
       author = {{Gibelyou}, Cameron and {Huterer}, Dragan},
        title = "{Dipoles in the sky}",
      journal = {\mnras},
         year = {2012},
        month = {dec},
       volume = {427},
       number = {3},
        pages = {1994-2021},
          doi = {10.1111/j.1365-2966.2012.22032.x},
archivePrefix = {arXiv},
       eprint = {1205.6476},
 primaryClass = {astro-ph.CO},
       adsurl = {https://ui.adsabs.harvard.edu/abs/2012MNRAS.427.1994G}
}

@ARTICLE{2013A&A...555A.117R,
       author = {{Rubart}, M. and {Schwarz}, D.~J.},
        title = "{Cosmic radio dipole from NVSS and WENSS}",
      journal = {\aap},
         year = {2013},
        month = {jul},
       volume = {555},
          eid = {A117},
        pages = {A117},
          doi = {10.1051/0004-6361/201321215},
archivePrefix = {arXiv},
       eprint = {1301.5559},
 primaryClass = {astro-ph.CO},
       adsurl = {https://ui.adsabs.harvard.edu/abs/2013A&A...555A.117R}
}

@ARTICLE{2010ApJ...712L..81K,
       author = {{Kashlinsky}, A. and {Atrio-Barandela}, F. and {Ebeling}, H. and {Edge}, A. and {Kocevski}, D.},
        title = "{A New Measurement of the Bulk Flow of X-Ray Luminous Clusters of Galaxies}",
      journal = {\apjl},
         year = {2010},
        month = {mar},
       volume = {712},
       number = {1},
        pages = {L81-L85},
          doi = {10.1088/2041-8205/712/1/L81},
archivePrefix = {arXiv},
       eprint = {0910.4958},
 primaryClass = {astro-ph.CO},
       adsurl = {https://ui.adsabs.harvard.edu/abs/2010ApJ...712L..81K}
}

@article{Das_2021,
   title={Explaining excess dipole in NVSS data using superhorizon perturbation},
   volume={2021},
   ISSN={1475-7516},
   url={http://dx.doi.org/10.1088/1475-7516/2021/07/035},
   DOI={10.1088/1475-7516/2021/07/035},
   number={07},
   journal={Journal of Cosmology and Astroparticle Physics},
   publisher={IOP Publishing},
   author={Das, Kaustav K. and Sankharva, Kishan and Jain, Pankaj},
   year={2021},
   month={jul},
    pages={035} 
}

@article{Zonca2019,
    doi = {10.21105/joss.01298},
    url = {https://doi.org/10.21105/joss.01298},
    year = {2019}, publisher = {The Open Journal},
    volume = {4},
    number = {35},
    pages = {1298},
    author = {Andrea Zonca and Leo P. Singer and Daniel Lenz and Martin Reinecke and Cyrille Rosset and Eric Hivon and Krzysztof M. Gorski},
    title = {healpy: equal area pixelization and spherical harmonics transforms for data on the sphere in Python},
    journal = {Journal of Open Source Software}
}

@article{Rath:2014cka,
    author = "Rath, Pranati K. and Aluri, Pavan K. and Jain, Pankaj",
    title = "{Relating the inhomogeneous power spectrum to the CMB hemispherical anisotropy}",
    eprint = "1403.2567",
    archivePrefix = "arXiv",
    primaryClass = "astro-ph.CO",
    doi = "10.1103/PhysRevD.91.023515",
    journal = "Phys. Rev. D",
    volume = "91",
    pages = "023515",
    year = "2015"
}

@book{Gorbunov:2011zzc,
    author = "Gorbunov, Dmitry S. and Rubakov, Valery A.",
    title = "{Introduction to the theory of the early universe: Cosmological perturbations and inflationary theory}",
    publisher = "World Scientific",
    doi = "10.1142/7873",
    year = "2011"
}

@article{Jain1998,
author = {Jain, Pankaj and Ralston, John P.},
title = {ANISOTROPY IN THE PROPAGATION OF RADIO POLARIZATIONS FROM COSMOLOGICALLY DISTANT GALAXIES},
journal = {Modern Physics Letters A},
volume = {14},
number = {06},
pages = {417-432},
year = {1999},
doi = {10.1142/S0217732399000481},
URL = {https://doi.org/10.1142/S0217732399000481}
}

@article{Ralston2004,
author = {{Ralston}, JOHN P. and {Jain}, PANKAJ},
title = {THE VIRGO ALIGNMENT PUZZLE IN PROPAGATION OF RADIATION ON COSMOLOGICAL SCALES},
journal = {International Journal of Modern Physics D},
volume = {13},
number = {09},
pages = {1857-1877},
year = {2004},
doi = {10.1142/S0218271804005948},
URL = {https://doi.org/10.1142/S0218271804005948}
}

@article{Wald_Robert_1983,
  title = {Asymptotic behavior of homogeneous cosmological models in the presence of a positive cosmological constant},
  author = {Wald, Robert M.},
  journal = {Phys. Rev. D},
  volume = {28},
  issue = {8},
  pages = {2118--2120},
  numpages = {0},
  year = {1983},
  month = {Oct},
  publisher = {American Physical Society},
  doi = {10.1103/PhysRevD.28.2118},
  url = {https://link.aps.org/doi/10.1103/PhysRevD.28.2118}
}

@article{PhysRevD.94.063531,
  title = {Cosmological power spectrum in a noncommutative spacetime},
  author = {Kothari, Rahul and Rath, Pranati K. and Jain, Pankaj},
  journal = {Phys. Rev. D},
  volume = {94},
  issue = {6},
  pages = {063531},
  numpages = {10},
  year = {2016},
  month = {Sep},
  publisher = {American Physical Society},
  doi = {10.1103/PhysRevD.94.063531},
  url = {https://link.aps.org/doi/10.1103/PhysRevD.94.063531}
}

@article{ refId0,
	author = {{Planck Collaboration} and {Akrami, Y.} and {Andersen, K. J.} and {Ashdown, M.} and {Baccigalupi, C.} and {Ballardini, M.} and {Banday, A. J.} and {Barreiro, R. B.} and {Bartolo, N.} and {Basak, S.} and {Benabed, K.} and {Bernard, J.-P.} and {Bersanelli, M.} and {Bielewicz, P.} and {Bond, J. R.} and {Borrill, J.} and {Burigana, C.} and {Butler, R. C.} and {Calabrese, E.} and {Casaponsa, B.} and {Chiang, H. C.} and {Colombo, L. P. L.} and {Combet, C.} and {Crill, B. P.} and {Cuttaia, F.} and {de Bernardis, P.} and {de Rosa, A.} and {de Zotti, G.} and {Delabrouille, J.} and {Di Valentino, E.} and {Diego, J. M.} and {Doré, O.} and {Douspis, M.} and {Dupac, X.} and {Eriksen, H. K.} and {Fernandez-Cobos, R.} and {Finelli, F.} and {Frailis, M.} and {Fraisse, A. A.} and {Franceschi, E.} and {Frolov, A.} and {Galeotta, S.} and {Galli, S.} and {Ganga, K.} and {Gerbino, M.} and {Ghosh, T.} and {González-Nuevo, J.} and {Górski, K. M.} and {Gruppuso, A.} and {Gudmundsson, J. E.} and {Handley, W.} and {Helou, G.} and {Herranz, D.} and {Hildebrandt, S. R.} and {Hivon, E.} and {Huang, Z.} and {Jaffe, A. H.} and {Jones, W. C.} and {Keihänen, E.} and {Keskitalo, R.} and {Kiiveri, K.} and {Kim, J.} and {Kisner, T. S.} and {Krachmalnicoff, N.} and {Kunz, M.} and {Kurki-Suonio, H.} and {Lasenby, A.} and {Lattanzi, M.} and {Lawrence, C. R.} and {Le Jeune, M.} and {Levrier, F.} and {Liguori, M.} and {Lilje, P. B.} and {Lilley, M.} and {Lindholm, V.} and {López-Caniego, M.} and {Lubin, P. M.} and {Macías-Pérez, J. F.} and {Maino, D.} and {Mandolesi, N.} and {Marcos-Caballero, A.} and {Maris, M.} and {Martin, P. G.} and {Martínez-González, E.} and {Matarrese, S.} and {Mauri, N.} and {McEwen, J. D.} and {Meinhold, P. R.} and {Mennella, A.} and {Migliaccio, M.} and {Mitra, S.} and {Molinari, D.} and {Montier, L.} and {Morgante, G.} and {Moss, A.} and {Natoli, P.} and {Paoletti, D.} and {Partridge, B.} and {Patanchon, G.} and {Pearson, D.} and {Pearson, T. J.} and {Perrotta, F.} and {Piacentini, F.} and {Polenta, G.} and {Rachen, J. P.} and {Reinecke, M.} and {Remazeilles, M.} and {Renzi, A.} and {Rocha, G.} and {Rosset, C.} and {Roudier, G.} and {Rubiño-Martín, J. A.} and {Ruiz-Granados, B.} and {Salvati, L.} and {Savelainen, M.} and {Scott, D.} and {Sirignano, C.} and {Sirri, G.} and {Spencer, L. D.} and {Suur-Uski, A.-S.} and {Svalheim, L. T.} and {Tauber, J. A.} and {Tavagnacco, D.} and {Tenti, M.} and {Terenzi, L.} and {Thommesen, H.} and {Toffolatti, L.} and {Tomasi, M.} and {Tristram, M.} and {Trombetti, T.} and {Valiviita, J.} and {Van Tent, B.} and {Vielva, P.} and {Villa, F.} and {Vittorio, N.} and {Wandelt, B. D.} and {Wehus, I. K.} and {Zacchei, A.} and {Zonca, A.}},
	title = {Planck intermediate results - LVII. Joint Planck LFI and HFI data processing},
	DOI= "10.1051/0004-6361/202038073",
	url= "https://doi.org/10.1051/0004-6361/202038073",
	journal = {A\&A},
	year = 2020,
	volume = 643,
	pages = "A42",
}

@article{Gorski_2005,
   title={HEALPix: A Framework for High‐Resolution Discretization and Fast Analysis of Data Distributed on the Sphere},
   volume={622},
   ISSN={1538-4357},
   url={http://dx.doi.org/10.1086/427976},
   DOI={10.1086/427976},
   number={2},
   journal={The Astrophysical Journal},
   publisher={American Astronomical Society},
   author={Gorski, K. M. and Hivon, E. and Banday, A. J. and Wandelt, B. D. and Hansen, F. K. and Reinecke, M. and Bartelmann, M.},
   year={2005},
   month=apr, pages={759–771} }

@article{Garc_a_Garc_a_2019,
   title={Disconnected pseudo-C$\ell$ covariances for projected large-scale structure data},
   volume={2019},
   ISSN={1475-7516},
   url={http://dx.doi.org/10.1088/1475-7516/2019/11/043},
   DOI={10.1088/1475-7516/2019/11/043},
   number={11},
   journal={Journal of Cosmology and Astroparticle Physics},
   publisher={IOP Publishing},
   author={García-García, Carlos and Alonso, David and Bellini, Emilio},
   year={2019},
   month=nov, pages={043–043} }

@article{2020,
   title={Planck2018 results: I. Overview and the cosmological legacy ofPlanck},
   volume={641},
   ISSN={1432-0746},
   url={http://dx.doi.org/10.1051/0004-6361/201833880},
   DOI={10.1051/0004-6361/201833880},
   journal={Astronomy \&amp; Astrophysics},
   publisher={EDP Sciences},
   author={Aghanim, N. and Akrami, Y. and Arroja, F. and Ashdown, M. and Aumont, J. and Baccigalupi, C. and Ballardini, M. and Banday, A. J. and Barreiro, R. B. and Bartolo, N. and Basak, S. and Battye, R. and Benabed, K. and Bernard, J.-P. and Bersanelli, M. and Bielewicz, P. and Bock, J. J. and Bond, J. R. and Borrill, J. and Bouchet, F. R. and Boulanger, F. and Bucher, M. and Burigana, C. and Butler, R. C. and Calabrese, E. and Cardoso, J.-F. and Carron, J. and Casaponsa, B. and Challinor, A. and Chiang, H. C. and Colombo, L. P. L. and Combet, C. and Contreras, D. and Crill, B. P. and Cuttaia, F. and de Bernardis, P. and de Zotti, G. and Delabrouille, J. and Delouis, J.-M. and Désert, F.-X. and Di Valentino, E. and Dickinson, C. and Diego, J. M. and Donzelli, S. and Doré, O. and Douspis, M. and Ducout, A. and Dupac, X. and Efstathiou, G. and Elsner, F. and Enßlin, T. A. and Eriksen, H. K. and Falgarone, E. and Fantaye, Y. and Fergusson, J. and Fernandez-Cobos, R. and Finelli, F. and Forastieri, F. and Frailis, M. and Franceschi, E. and Frolov, A. and Galeotta, S. and Galli, S. and Ganga, K. and Génova-Santos, R. T. and Gerbino, M. and Ghosh, T. and González-Nuevo, J. and Górski, K. M. and Gratton, S. and Gruppuso, A. and Gudmundsson, J. E. and Hamann, J. and Handley, W. and Hansen, F. K. and Helou, G. and Herranz, D. and Hildebrandt, S. R. and Hivon, E. and Huang, Z. and Jaffe, A. H. and Jones, W. C. and Karakci, A. and Keihänen, E. and Keskitalo, R. and Kiiveri, K. and Kim, J. and Kisner, T. S. and Knox, L. and Krachmalnicoff, N. and Kunz, M. and Kurki-Suonio, H. and Lagache, G. and Lamarre, J.-M. and Langer, M. and Lasenby, A. and Lattanzi, M. and Lawrence, C. R. and Le Jeune, M. and Leahy, J. P. and Lesgourgues, J. and Levrier, F. and Lewis, A. and Liguori, M. and Lilje, P. B. and Lilley, M. and Lindholm, V. and López-Caniego, M. and Lubin, P. M. and Ma, Y.-Z. and Macías-Pérez, J. F. and Maggio, G. and Maino, D. and Mandolesi, N. and Mangilli, A. and Marcos-Caballero, A. and Maris, M. and Martin, P. G. and Martinelli, M. and Martínez-González, E. and Matarrese, S. and Mauri, N. and McEwen, J. D. and Meerburg, P. D. and Meinhold, P. R. and Melchiorri, A. and Mennella, A. and Migliaccio, M. and Millea, M. and Mitra, S. and Miville-Deschênes, M.-A. and Molinari, D. and Moneti, A. and Montier, L. and Morgante, G. and Moss, A. and Mottet, S. and Münchmeyer, M. and Natoli, P. and Nørgaard-Nielsen, H. U. and Oxborrow, C. A. and Pagano, L. and Paoletti, D. and Partridge, B. and Patanchon, G. and Pearson, T. J. and Peel, M. and Peiris, H. V. and Perrotta, F. and Pettorino, V. and Piacentini, F. and Polastri, L. and Polenta, G. and Puget, J.-L. and Rachen, J. P. and Reinecke, M. and Remazeilles, M. and Renault, C. and Renzi, A. and Rocha, G. and Rosset, C. and Roudier, G. and Rubiño-Martín, J. A. and Ruiz-Granados, B. and Salvati, L. and Sandri, M. and Savelainen, M. and Scott, D. and Shellard, E. P. S. and Shiraishi, M. and Sirignano, C. and Sirri, G. and Spencer, L. D. and Sunyaev, R. and Suur-Uski, A.-S. and Tauber, J. A. and Tavagnacco, D. and Tenti, M. and Terenzi, L. and Toffolatti, L. and Tomasi, M. and Trombetti, T. and Valiviita, J. and Van Tent, B. and Vibert, L. and Vielva, P. and Villa, F. and Vittorio, N. and Wandelt, B. D. and Wehus, I. K. and White, M. and White, S. D. M. and Zacchei, A. and Zonca, A.},
   year={2020},
   month=sep, pages={A1} }

@article{Alonso_2019,
   title={A unified pseudo-C$\ell$ framework},
   volume={484},
   ISSN={1365-2966},
   url={http://dx.doi.org/10.1093/mnras/stz093},
   DOI={10.1093/mnras/stz093},
   number={3},
   journal={Monthly Notices of the Royal Astronomical Society},
   publisher={Oxford University Press (OUP)},
   author={Alonso, David and Sanchez, Javier and Slosar, Anže},
   year={2019},
   month=jan, pages={4127–4151} }

@article{Lewis_2000,
   title={Efficient Computation of Cosmic Microwave Background Anisotropies in Closed Friedmann‐Robertson‐Walker Models},
   volume={538},
   ISSN={1538-4357},
   url={http://dx.doi.org/10.1086/309179},
   DOI={10.1086/309179},
   number={2},
   journal={The Astrophysical Journal},
   publisher={American Astronomical Society},
   author={Lewis, Antony and Challinor, Anthony and Lasenby, Anthony},
   year={2000},
   month=aug, pages={473–476} }

@article{Howlett_2012,
   title={CMB power spectrum parameter degeneracies in the era of precision cosmology},
   volume={2012},
   ISSN={1475-7516},
   url={http://dx.doi.org/10.1088/1475-7516/2012/04/027},
   DOI={10.1088/1475-7516/2012/04/027},
   number={04},
   journal={Journal of Cosmology and Astroparticle Physics},
   publisher={IOP Publishing},
   author={Howlett, Cullan and Lewis, Antony and Hall, Alex and Challinor, Anthony},
   year={2012},
   month=apr, pages={027–027} }

@article{Tristram_2024,
   title={Cosmological parameters derived from the final Planck data release (PR4)},
   volume={682},
   ISSN={1432-0746},
   url={http://dx.doi.org/10.1051/0004-6361/202348015},
   DOI={10.1051/0004-6361/202348015},
   journal={Astronomy \&amp; Astrophysics},
   publisher={EDP Sciences},
   author={Tristram, M. and Banday, A. J. and Douspis, M. and Garrido, X. and Górski, K. M. and Henrot-Versillé, S. and Hergt, L. T. and Ilić, S. and Keskitalo, R. and Lagache, G. and Lawrence, C. R. and Partridge, B. and Scott, D.},
   year={2024},
   month=jan, pages={A37} }

@article{Virtanen_2020,
   title={SciPy 1.0: fundamental algorithms for scientific computing in Python},
   volume={17},
   ISSN={1548-7105},
   url={http://dx.doi.org/10.1038/s41592-019-0686-2},
   DOI={10.1038/s41592-019-0686-2},
   number={3},
   journal={Nature Methods},
   publisher={Springer Science and Business Media LLC},
   author={Virtanen, Pauli and Gommers, Ralf and Oliphant, Travis E. and Haberland, Matt and Reddy, Tyler and Cournapeau, David and Burovski, Evgeni and Peterson, Pearu and Weckesser, Warren and Bright, Jonathan and van der Walt, Stéfan J. and Brett, Matthew and Wilson, Joshua and Millman, K. Jarrod and Mayorov, Nikolay and Nelson, Andrew R. J. and Jones, Eric and Kern, Robert and Larson, Eric and Carey, C J and Polat, İlhan and Feng, Yu and Moore, Eric W. and VanderPlas, Jake and Laxalde, Denis and Perktold, Josef and Cimrman, Robert and Henriksen, Ian and Quintero, E. A. and Harris, Charles R. and Archibald, Anne M. and Ribeiro, Antônio H. and Pedregosa, Fabian and van Mulbregt, Paul and Vijaykumar, Aditya and Bardelli, Alessandro Pietro and Rothberg, Alex and Hilboll, Andreas and Kloeckner, Andreas and Scopatz, Anthony and Lee, Antony and Rokem, Ariel and Woods, C. Nathan and Fulton, Chad and Masson, Charles and Häggström, Christian and Fitzgerald, Clark and Nicholson, David A. and Hagen, David R. and Pasechnik, Dmitrii V. and Olivetti, Emanuele and Martin, Eric and Wieser, Eric and Silva, Fabrice and Lenders, Felix and Wilhelm, Florian and Young, G. and Price, Gavin A. and Ingold, Gert-Ludwig and Allen, Gregory E. and Lee, Gregory R. and Audren, Hervé and Probst, Irvin and Dietrich, Jörg P. and Silterra, Jacob and Webber, James T and Slavič, Janko and Nothman, Joel and Buchner, Johannes and Kulick, Johannes and Schönberger, Johannes L. and de Miranda Cardoso, José Vinícius and Reimer, Joscha and Harrington, Joseph and Rodríguez, Juan Luis Cano and Nunez-Iglesias, Juan and Kuczynski, Justin and Tritz, Kevin and Thoma, Martin and Newville, Matthew and Kümmerer, Matthias and Bolingbroke, Maximilian and Tartre, Michael and Pak, Mikhail and Smith, Nathaniel J. and Nowaczyk, Nikolai and Shebanov, Nikolay and Pavlyk, Oleksandr and Brodtkorb, Per A. and Lee, Perry and McGibbon, Robert T. and Feldbauer, Roman and Lewis, Sam and Tygier, Sam and Sievert, Scott and Vigna, Sebastiano and Peterson, Stefan and More, Surhud and Pudlik, Tadeusz and Oshima, Takuya and Pingel, Thomas J. and Robitaille, Thomas P. and Spura, Thomas and Jones, Thouis R. and Cera, Tim and Leslie, Tim and Zito, Tiziano and Krauss, Tom and Upadhyay, Utkarsh and Halchenko, Yaroslav O. and Vázquez-Baeza, Yoshiki},
   year={2020},
   month=feb, pages={261–272} }

@article{article,
author = {Hunter, John},
year = {2007},
month = {06},
pages = {90-95},
title = {Matplotlib: A 2D Graphics Environment},
volume = {9},
journal = {Computing in Science \& Engineering},
doi = {10.1109/MCSE.2007.55}
}

@Article{         harris2020array,
 title         = {Array programming with {NumPy}},
 author        = {Charles R. Harris and K. Jarrod Millman and St{\'{e}}fan J.
                 van der Walt and Ralf Gommers and Pauli Virtanen and David
                 Cournapeau and Eric Wieser and Julian Taylor and Sebastian
                 Berg and Nathaniel J. Smith and Robert Kern and Matti Picus
                 and Stephan Hoyer and Marten H. van Kerkwijk and Matthew
                 Brett and Allan Haldane and Jaime Fern{\'{a}}ndez del
                 R{\'{i}}o and Mark Wiebe and Pearu Peterson and Pierre
                 G{\'{e}}rard-Marchant and Kevin Sheppard and Tyler Reddy and
                 Warren Weckesser and Hameer Abbasi and Christoph Gohlke and
                 Travis E. Oliphant},
 year          = {2020},
 month         = sep,
 journal       = {Nature},
 volume        = {585},
 number        = {7825},
 pages         = {357--362},
 doi           = {10.1038/s41586-020-2649-2},
 publisher     = {Springer Science and Business Media {LLC}},
 url           = {https://doi.org/10.1038/s41586-020-2649-2}
}

@ARTICLE{2022ApJ...935..167A,
       author = {{Astropy Collaboration} and {Price-Whelan}, Adrian M. and {Lim}, Pey Lian and {Earl}, Nicholas and {Starkman}, Nathaniel and {Bradley}, Larry and {Shupe}, David L. and {Patil}, Aarya A. and {Corrales}, Lia and {Brasseur}, C.~E. and {N{\"o}the}, Maximilian and {Donath}, Axel and {Tollerud}, Erik and {Morris}, Brett M. and {Ginsburg}, Adam and {Vaher}, Eero and {Weaver}, Benjamin A. and {Tocknell}, James and {Jamieson}, William and {van Kerkwijk}, Marten H. and {Robitaille}, Thomas P. and {Merry}, Bruce and {Bachetti}, Matteo and {G{\"u}nther}, H. Moritz and {Aldcroft}, Thomas L. and {Alvarado-Montes}, Jaime A. and {Archibald}, Anne M. and {B{\'o}di}, Attila and {Bapat}, Shreyas and {Barentsen}, Geert and {Baz{\'a}n}, Juanjo and {Biswas}, Manish and {Boquien}, M{\'e}d{\'e}ric and {Burke}, D.~J. and {Cara}, Daria and {Cara}, Mihai and {Conroy}, Kyle E. and {Conseil}, Simon and {Craig}, Matthew W. and {Cross}, Robert M. and {Cruz}, Kelle L. and {D'Eugenio}, Francesco and {Dencheva}, Nadia and {Devillepoix}, Hadrien A.~R. and {Dietrich}, J{\"o}rg P. and {Eigenbrot}, Arthur Davis and {Erben}, Thomas and {Ferreira}, Leonardo and {Foreman-Mackey}, Daniel and {Fox}, Ryan and {Freij}, Nabil and {Garg}, Suyog and {Geda}, Robel and {Glattly}, Lauren and {Gondhalekar}, Yash and {Gordon}, Karl D. and {Grant}, David and {Greenfield}, Perry and {Groener}, Austen M. and {Guest}, Steve and {Gurovich}, Sebastian and {Handberg}, Rasmus and {Hart}, Akeem and {Hatfield-Dodds}, Zac and {Homeier}, Derek and {Hosseinzadeh}, Griffin and {Jenness}, Tim and {Jones}, Craig K. and {Joseph}, Prajwel and {Kalmbach}, J. Bryce and {Karamehmetoglu}, Emir and {Ka{\l}uszy{\'n}ski}, Miko{\l}aj and {Kelley}, Michael S.~P. and {Kern}, Nicholas and {Kerzendorf}, Wolfgang E. and {Koch}, Eric W. and {Kulumani}, Shankar and {Lee}, Antony and {Ly}, Chun and {Ma}, Zhiyuan and {MacBride}, Conor and {Maljaars}, Jakob M. and {Muna}, Demitri and {Murphy}, N.~A. and {Norman}, Henrik and {O'Steen}, Richard and {Oman}, Kyle A. and {Pacifici}, Camilla and {Pascual}, Sergio and {Pascual-Granado}, J. and {Patil}, Rohit R. and {Perren}, Gabriel I. and {Pickering}, Timothy E. and {Rastogi}, Tanuj and {Roulston}, Benjamin R. and {Ryan}, Daniel F. and {Rykoff}, Eli S. and {Sabater}, Jose and {Sakurikar}, Parikshit and {Salgado}, Jes{\'u}s and {Sanghi}, Aniket and {Saunders}, Nicholas and {Savchenko}, Volodymyr and {Schwardt}, Ludwig and {Seifert-Eckert}, Michael and {Shih}, Albert Y. and {Jain}, Anany Shrey and {Shukla}, Gyanendra and {Sick}, Jonathan and {Simpson}, Chris and {Singanamalla}, Sudheesh and {Singer}, Leo P. and {Singhal}, Jaladh and {Sinha}, Manodeep and {Sip{\H{o}}cz}, Brigitta M. and {Spitler}, Lee R. and {Stansby}, David and {Streicher}, Ole and {{\v{S}}umak}, Jani and {Swinbank}, John D. and {Taranu}, Dan S. and {Tewary}, Nikita and {Tremblay}, Grant R. and {de Val-Borro}, Miguel and {Van Kooten}, Samuel J. and {Vasovi{\'c}}, Zlatan and {Verma}, Shresth and {de Miranda Cardoso}, Jos{\'e} Vin{\'\i}cius and {Williams}, Peter K.~G. and {Wilson}, Tom J. and {Winkel}, Benjamin and {Wood-Vasey}, W.~M. and {Xue}, Rui and {Yoachim}, Peter and {Zhang}, Chen and {Zonca}, Andrea and {Astropy Project Contributors}},
        title = "{The Astropy Project: Sustaining and Growing a Community-oriented Open-source Project and the Latest Major Release (v5.0) of the Core Package}",
      journal = {\apj},
         year = 2022,
        month = aug,
       volume = {935},
       number = {2},
          eid = {167},
        pages = {167},
          doi = {10.3847/1538-4357/ac7c74},
archivePrefix = {arXiv},
       eprint = {2206.14220},
 primaryClass = {astro-ph.IM},
       adsurl = {https://ui.adsabs.harvard.edu/abs/2022ApJ...935..167A}
}

@article{Pranati_K_Rath_2013,
doi = {10.1088/1475-7516/2013/12/014},
url = {https://dx.doi.org/10.1088/1475-7516/2013/12/014},
year = {2013},
month = {dec},
publisher = {},
volume = {2013},
number = {12},
pages = {014},
author = {Pranati K. Rath and Pankaj Jain},
title = {Testing the dipole modulation model in CMBR},
journal = {Journal of Cosmology and Astroparticle Physics}
}

@article{ refId1,
	author = {{Planck Collaboration} and {Akrami, Y.} and {Ashdown, M.} and {Aumont, J.} and {Baccigalupi, C.} and {Ballardini, M.} and {Banday, A. J.} and {Barreiro, R. B.} and {Bartolo, N.} and {Basak, S.} and {Benabed, K.} and {Bersanelli, M.} and {Bielewicz, P.} and {Bock, J. J.} and {Bond, J. R.} and {Borrill, J.} and {Bouchet, F. R.} and {Boulanger, F.} and {Bucher, M.} and {Burigana, C.} and {Butler, R. C.} and {Calabrese, E.} and {Cardoso, J.-F.} and {Casaponsa, B.} and {Chiang, H. C.} and {Colombo, L. P. L.} and {Combet, C.} and {Contreras, D.} and {Crill, B. P.} and {de Bernardis, P.} and {de Zotti, G.} and {Delabrouille, J.} and {Delouis, J.-M.} and {Di Valentino, E.} and {Diego, J. M.} and {Doré, O.} and {Douspis, M.} and {Ducout, A.} and {Dupac, X.} and {Efstathiou, G.} and {Elsner, F.} and {Enßlin, T. A.} and {Eriksen, H. K.} and {Fantaye, Y.} and {Fernandez-Cobos, R.} and {Finelli, F.} and {Frailis, M.} and {Fraisse, A. A.} and {Franceschi, E.} and {Frolov, A.} and {Galeotta, S.} and {Galli, S.} and {Ganga, K.} and {Génova-Santos, R. T.} and {Gerbino, M.} and {Ghosh, T.} and {González-Nuevo, J.} and {Górski, K. M.} and {Gruppuso, A.} and {Gudmundsson, J. E.} and {Hamann, J.} and {Handley, W.} and {Hansen, F. K.} and {Herranz, D.} and {Hivon, E.} and {Huang, Z.} and {Jaffe, A. H.} and {Jones, W. C.} and {Keihänen, E.} and {Keskitalo, R.} and {Kiiveri, K.} and {Kim, J.} and {Krachmalnicoff, N.} and {Kunz, M.} and {Kurki-Suonio, H.} and {Lagache, G.} and {Lamarre, J.-M.} and {Lasenby, A.} and {Lattanzi, M.} and {Lawrence, C. R.} and {Le Jeune, M.} and {Levrier, F.} and {Liguori, M.} and {Lilje, P. B.} and {Lindholm, V.} and {López-Caniego, M.} and {Ma, Y.-Z.} and {Macías-Pérez, J. F.} and {Maggio, G.} and {Maino, D.} and {Mandolesi, N.} and {Mangilli, A.} and {Marcos-Caballero, A.} and {Maris, M.} and {Martin, P. G.} and {Martínez-González, E.} and {Matarrese, S.} and {Mauri, N.} and {McEwen, J. D.} and {Meinhold, P. R.} and {Mennella, A.} and {Migliaccio, M.} and {Miville-Deschênes, M.-A.} and {Molinari, D.} and {Moneti, A.} and {Montier, L.} and {Morgante, G.} and {Moss, A.} and {Natoli, P.} and {Pagano, L.} and {Paoletti, D.} and {Partridge, B.} and {Perrotta, F.} and {Pettorino, V.} and {Piacentini, F.} and {Polenta, G.} and {Puget, J.-L.} and {Rachen, J. P.} and {Reinecke, M.} and {Remazeilles, M.} and {Renzi, A.} and {Rocha, G.} and {Rosset, C.} and {Roudier, G.} and {Rubiño-Martín, J. A.} and {Ruiz-Granados, B.} and {Salvati, L.} and {Savelainen, M.} and {Scott, D.} and {Shellard, E. P. S.} and {Sirignano, C.} and {Sunyaev, R.} and {Suur-Uski, A.-S.} and {Tauber, J. A.} and {Tavagnacco, D.} and {Tenti, M.} and {Toffolatti, L.} and {Tomasi, M.} and {Trombetti, T.} and {Valenziano, L.} and {Valiviita, J.} and {Van Tent, B.} and {Vielva, P.} and {Villa, F.} and {Vittorio, N.} and {Wandelt, B. D.} and {Wehus, I. K.} and {Zacchei, A.} and {Zibin, J. P.} and {Zonca, A.}},
	title = {Planck 2018 results - VII. Isotropy and statistics of the CMB},
	DOI= "10.1051/0004-6361/201935201",
	url= "https://doi.org/10.1051/0004-6361/201935201",
	journal = {A\&A},
	year = 2020,
	volume = 641,
	pages = "A7",
}

@article{Agullo:2021oqk,
    author = "Agullo, Ivan and Kranas, Dimitrios and Sreenath, V.",
    title = "{Anomalies in the Cosmic Microwave Background and Their Non-Gaussian Origin in Loop Quantum Cosmology}",
    eprint = "2105.12993",
    archivePrefix = "arXiv",
    primaryClass = "gr-qc",
    doi = "10.3389/fspas.2021.703845",
    journal = "Front. Astron. Space Sci.",
    volume = "8",
    pages = "703845",
    year = "2021"
}

@article{10.1093/mnras/stu932,
    author = {Russell, Esra and Kılınç, Can Battal and Pashaev, Oktay K.},
    title = {Bianchi I model: an alternative way to model the present-day Universe},
    journal = {Monthly Notices of the Royal Astronomical Society},
    volume = {442},
    number = {3},
    pages = {2331-2341},
    year = {2014},
    month = {06},
    issn = {0035-8711},
    doi = {10.1093/mnras/stu932},
    url = {https://doi.org/10.1093/mnras/stu932}}

@article{10.1111/j.1365-2966.2007.12221.x,
    author = {Pontzen, Andrew and Challinor, Anthony},
    title = {Bianchi model CMB polarization and its implications for CMB anomalies},
    journal = {Monthly Notices of the Royal Astronomical Society},
    volume = {380},
    number = {4},
    pages = {1387-1398},
    year = {2007},
    month = {09},
    issn = {0035-8711},
    doi = {10.1111/j.1365-2966.2007.12221.x},
    url = {https://doi.org/10.1111/j.1365-2966.2007.12221.x},
    }

@article{ refId2,
	author = {{Jung, Gabriel} and {Aghanim, Nabila} and {Sorce, Jenny G.} and {Seidel, Benjamin} and {Dolag, Klaus} and {Douspis, Marian}},
	title = {Revisiting the CMB large-scale anomalies: The impact of the Sunyaev-Zeldovich signal from the Local Universe},
	DOI= "10.1051/0004-6361/202451238",
	url= "https://doi.org/10.1051/0004-6361/202451238",
	journal = {A\&A},
	year = 2024,
	volume = 692,
	pages = "A180",
}

@article{ refId3,
	author = {{Hansen, Frode K.} and {Boero, Ezequiel F.} and {Luparello, Heliana E.} and {Garcia Lambas, Diego}},
	title = {A possible common explanation for several cosmic microwave background (CMB) anomalies: A strong impact of nearby galaxies on observed large-scale CMB fluctuations},
	DOI= "10.1051/0004-6361/202346779",
	url= "https://doi.org/10.1051/0004-6361/202346779",
	journal = {A\&A},
	year = 2023,
	volume = 675,
	pages = "L7",
}

@article{Bolejko_2011,
   title={Inhomogeneous cosmological models: exact solutions and their applications},
   volume={28},
   ISSN={1361-6382},
   url={http://dx.doi.org/10.1088/0264-9381/28/16/164002},
   DOI={10.1088/0264-9381/28/16/164002},
   number={16},
   journal={Classical and Quantum Gravity},
   publisher={IOP Publishing},
   author={Bolejko, Krzysztof and Célérier, Marie-Noëlle and Krasiński, Andrzej},
   year={2011},
   month=aug, pages={164002} }

@article{PhysRevLett.111.251302,
  title = {Stringent Restriction from the Growth of Large-Scale Structure on Apparent Acceleration in Inhomogeneous Cosmological Models},
  author = {Ishak, Mustapha and Peel, Austin and Troxel, M. A.},
  journal = {Phys. Rev. Lett.},
  volume = {111},
  issue = {25},
  pages = {251302},
  numpages = {5},
  year = {2013},
  month = {Dec},
  publisher = {American Physical Society},
  doi = {10.1103/PhysRevLett.111.251302},
  url = {https://link.aps.org/doi/10.1103/PhysRevLett.111.251302}
}

@article{Nistane_2019,
   title={CMB sky for an off-center observer in a local void. Part I. Framework for forecasts},
   volume={2019},
   ISSN={1475-7516},
   url={http://dx.doi.org/10.1088/1475-7516/2019/12/038},
   DOI={10.1088/1475-7516/2019/12/038},
   number={12},
   journal={Journal of Cosmology and Astroparticle Physics},
   publisher={IOP Publishing},
   author={Nistane, Viraj and Cusin, Giulia and Kunz, Martin},
   year={2019},
   month=dec, pages={038–038} }

@article{E_Akofor_2008,
doi = {10.1088/1126-6708/2008/05/092},
url = {https://dx.doi.org/10.1088/1126-6708/2008/05/092},
year = {2008},
month = {may},
publisher = {},
volume = {2008},
number = {05},
pages = {092},
author = {E. Akofor and A.P. Balachandran and S.G. Jo and A. Joseph and B.A. Qureshi},
title = {Direction-dependent CMB power spectrum and statistical anisotropy from noncommutative geometry},
journal = {Journal of High Energy Physics}
}

@article{Lyth:2006gd,
    author = "Lyth, David H.",
    title = "{Non-gaussianity and cosmic uncertainty in curvaton-type models}",
    eprint = "astro-ph/0602285",
    archivePrefix = "arXiv",
    doi = "10.1088/1475-7516/2006/06/015",
    journal = "JCAP",
    volume = "06",
    pages = "015",
    year = "2006"
}

@article{PhysRevD.90.023523,
  title = {Comprehensive analysis of the simplest curvaton model},
  author = {Byrnes, Christian T. and Cort\^es, Marina and Liddle, Andrew R.},
  journal = {Phys. Rev. D},
  volume = {90},
  issue = {2},
  pages = {023523},
  numpages = {8},
  year = {2014},
  month = {Jul},
  publisher = {American Physical Society},
  doi = {10.1103/PhysRevD.90.023523},
  url = {https://link.aps.org/doi/10.1103/PhysRevD.90.023523}
}

@article{Hoftuft_2009,
doi = {10.1088/0004-637X/699/2/985},
url = {https://dx.doi.org/10.1088/0004-637X/699/2/985},
year = {2009},
month = {jun},
publisher = {The American Astronomical Society},
volume = {699},
number = {2},
pages = {985},
author = {Hoftuft, J. and Eriksen, H. K. and Banday, A. J. and Górski, K. M. and Hansen, F. K. and Lilje, P. B.},
title = {INCREASING EVIDENCE FOR HEMISPHERICAL POWER ASYMMETRY IN THE FIVE-YEAR WMAP DATA},
journal = {The Astrophysical Journal}
}

@article{2014,
   title={Planck2013 results. XXIII. Isotropy and statistics of the CMB},
   volume={571},
   ISSN={1432-0746},
   url={http://dx.doi.org/10.1051/0004-6361/201321534},
   DOI={10.1051/0004-6361/201321534},
   journal={Astronomy \&amp; Astrophysics},
   publisher={EDP Sciences},
   author={Ade, P. A. R. and Aghanim, N. and Armitage-Caplan, C. and Arnaud, M. and Ashdown, M. and Atrio-Barandela, F. and Aumont, J. and Baccigalupi, C. and Banday, A. J. and Barreiro, R. B. and Bartlett, J. G. and Bartolo, N. and Battaner, E. and Battye, R. and Benabed, K. and Benoît, A. and Benoit-Lévy, A. and Bernard, J.-P. and Bersanelli, M. and Bielewicz, P. and Bobin, J. and Bock, J. J. and Bonaldi, A. and Bonavera, L. and Bond, J. R. and Borrill, J. and Bouchet, F. R. and Bridges, M. and Bucher, M. and Burigana, C. and Butler, R. C. and Cardoso, J.-F. and Catalano, A. and Challinor, A. and Chamballu, A. and Chary, R.-R. and Chiang, H. C. and Chiang, L.-Y and Christensen, P. R. and Church, S. and Clements, D. L. and Colombi, S. and Colombo, L. P. L. and Couchot, F. and Coulais, A. and Crill, B. P. and Cruz, M. and Curto, A. and Cuttaia, F. and Danese, L. and Davies, R. D. and Davis, R. J. and de Bernardis, P. and de Rosa, A. and de Zotti, G. and Delabrouille, J. and Delouis, J.-M. and Désert, F.-X. and Diego, J. M. and Dole, H. and Donzelli, S. and Doré, O. and Douspis, M. and Ducout, A. and Dupac, X. and Efstathiou, G. and Elsner, F. and Enßlin, T. A. and Eriksen, H. K. and Fantaye, Y. and Fergusson, J. and Finelli, F. and Forni, O. and Frailis, M. and Franceschi, E. and Frommert, M. and Galeotta, S. and Ganga, K. and Giard, M. and Giardino, G. and Giraud-Héraud, Y. and González-Nuevo, J. and Górski, K. M. and Gratton, S. and Gregorio, A. and Gruppuso, A. and Hansen, F. K. and Hansen, M. and Hanson, D. and Harrison, D. L. and Helou, G. and Henrot-Versillé, S. and Hernández-Monteagudo, C. and Herranz, D. and Hildebrandt, S. R. and Hivon, E. and Hobson, M. and Holmes, W. A. and Hornstrup, A. and Hovest, W. and Huffenberger, K. M. and Jaffe, A. H. and Jaffe, T. R. and Jones, W. C. and Juvela, M. and Keihänen, E. and Keskitalo, R. and Kim, J. and Kisner, T. S. and Knoche, J. and Knox, L. and Kunz, M. and Kurki-Suonio, H. and Lagache, G. and Lähteenmäki, A. and Lamarre, J.-M. and Lasenby, A. and Laureijs, R. J. and Lawrence, C. R. and Leahy, J. P. and Leonardi, R. and Leroy, C. and Lesgourgues, J. and Liguori, M. and Lilje, P. B. and Linden-Vørnle, M. and López-Caniego, M. and Lubin, P. M. and Macías-Pérez, J. F. and Maffei, B. and Maino, D. and Mandolesi, N. and Mangilli, A. and Marinucci, D. and Maris, M. and Marshall, D. J. and Martin, P. G. and Martínez-González, E. and Masi, S. and Massardi, M. and Matarrese, S. and Matthai, F. and Mazzotta, P. and McEwen, J. D. and Meinhold, P. R. and Melchiorri, A. and Mendes, L. and Mennella, A. and Migliaccio, M. and Mikkelsen, K. and Mitra, S. and Miville-Deschênes, M.-A. and Molinari, D. and Moneti, A. and Montier, L. and Morgante, G. and Mortlock, D. and Moss, A. and Munshi, D. and Murphy, J. A. and Naselsky, P. and Nati, F. and Natoli, P. and Netterfield, C. B. and Nørgaard-Nielsen, H. U. and Noviello, F. and Novikov, D. and Novikov, I. and Osborne, S. and Oxborrow, C. A. and Paci, F. and Pagano, L. and Pajot, F. and Paoletti, D. and Pasian, F. and Patanchon, G. and Peiris, H. V. and Perdereau, O. and Perotto, L. and Perrotta, F. and Piacentini, F. and Piat, M. and Pierpaoli, E. and Pietrobon, D. and Plaszczynski, S. and Pogosyan, D. and Pointecouteau, E. and Polenta, G. and Ponthieu, N. and Popa, L. and Poutanen, T. and Pratt, G. W. and Prézeau, G. and Prunet, S. and Puget, J.-L. and Rachen, J. P. and Racine, B. and Räth, C. and Rebolo, R. and Reinecke, M. and Remazeilles, M. and Renault, C. and Renzi, A. and Ricciardi, S. and Riller, T. and Ristorcelli, I. and Rocha, G. and Rosset, C. and Rotti, A. and Roudier, G. and Rubiño-Martín, J. A. and Ruiz-Granados, B. and Rusholme, B. and Sandri, M. and Santos, D. and Savini, G. and Scott, D. and Seiffert, M. D. and Shellard, E. P. S. and Souradeep, T. and Spencer, L. D. and Starck, J.-L. and Stolyarov, V. and Stompor, R. and Sudiwala, R. and Sureau, F. and Sutter, P. and Sutton, D. and Suur-Uski, A.-S. and Sygnet, J.-F. and Tauber, J. A. and Tavagnacco, D. and Terenzi, L. and Toffolatti, L. and Tomasi, M. and Tristram, M. and Tucci, M. and Tuovinen, J. and Türler, M. and Valenziano, L. and Valiviita, J. and Van Tent, B. and Varis, J. and Vielva, P. and Villa, F. and Vittorio, N. and Wade, L. A. and Wandelt, B. D. and Wehus, I. K. and White, M. and Wilkinson, A. and Yvon, D. and Zacchei, A. and Zonca, A.},
   year={2014},
   month=oct, pages={A23} }

@article{PhysRevD.101.123508,
  title = {Towards testing CMB anomalies using the kinetic and polarized Sunyaev-Zel'dovich effects},
  author = {Cayuso, Juan I. and Johnson, Matthew C.},
  journal = {Phys. Rev. D},
  volume = {101},
  issue = {12},
  pages = {123508},
  numpages = {16},
  year = {2020},
  month = {Jun},
  publisher = {American Physical Society},
  doi = {10.1103/PhysRevD.101.123508},
  url = {https://link.aps.org/doi/10.1103/PhysRevD.101.123508}
}

@book{Dodelson:2020bqr,
    author = "Dodelson, Scott and Schmidt, Fabian",
    title = "{Modern Cosmology}",
    doi = "10.1016/C2017-0-01943-2",
    publisher = "Academic Press",
    year = "2020"
}

@article{PhysRevD.80.063004,
  title = {Estimators for CMB statistical anisotropy},
  author = {Hanson, Duncan and Lewis, Antony},
  journal = {Phys. Rev. D},
  volume = {80},
  issue = {6},
  pages = {063004},
  numpages = {15},
  year = {2009},
  month = {Sep},
  publisher = {American Physical Society},
  doi = {10.1103/PhysRevD.80.063004},
  url = {https://link.aps.org/doi/10.1103/PhysRevD.80.063004}
}

@misc{sanyal2025reassessmentlvemethodhemispherical,
      title={A reassessment of LVE method and hemispherical power asymmetry in CMB temperature data from Planck PR4}, 
      author={Sanjeev Sanyal and Sanjeet K. Patel and Pavan K. Aluri and Arman Shafieloo},
      year={2025},
      eprint={2411.15786},
      archivePrefix={arXiv},
      primaryClass={astro-ph.CO},
      url={https://arxiv.org/abs/2411.15786}, 
}

@article{Gimeno-Amo_2023,
doi = {10.1088/1475-7516/2023/12/029},
url = {https://doi.org/10.1088/1475-7516/2023/12/029},
year = {2023},
month = {dec},
publisher = {IOP Publishing},
volume = {2023},
number = {12},
pages = {029},
author = {Gimeno-Amo, C. and Barreiro, R.B. and Martínez-González, E. and Marcos-Caballero, A.},
title = {Hemispherical power asymmetry in intensity and polarization for Planck PR4 data},
journal = {Journal of Cosmology and Astroparticle Physics}
}

@article{PhysRevD.75.083502,
  title = {Imprints of a primordial preferred direction on the microwave background},
  author = {Ackerman, Lotty and Carroll, Sean M. and Wise, Mark B.},
  journal = {Phys. Rev. D},
  volume = {75},
  issue = {8},
  pages = {083502},
  numpages = {7},
  year = {2007},
  month = {Apr},
  publisher = {American Physical Society},
  doi = {10.1103/PhysRevD.75.083502},
  url = {https://link.aps.org/doi/10.1103/PhysRevD.75.083502}
}

@article{Eriksen_2007,
doi = {10.1086/518091},
url = {https://doi.org/10.1086/518091},
year = {2007},
month = {apr},
publisher = {},
volume = {660},
number = {2},
pages = {L81},
author = {Eriksen, H. K. and Banday, A. J. and Górski, K. M. and Hansen, F. K. and Lilje, P. B.},
title = {Hemispherical Power Asymmetry in the Third-Year Wilkinson Microwave Anisotropy Probe Sky Maps},
journal = {The Astrophysical Journal}
}

@article{10.1111/j.1365-2966.2007.12707.x,
    author = {Gordon, Christopher and Trotta, Roberto},
    title = {Bayesian calibrated significance levels applied to the spectral tilt and hemispherical asymmetry},
    journal = {Monthly Notices of the Royal Astronomical Society},
    volume = {382},
    number = {4},
    pages = {1859-1863},
    year = {2007},
    month = {12},
    issn = {0035-8711},
    doi = {10.1111/j.1365-2966.2007.12707.x},
    url = {https://doi.org/10.1111/j.1365-2966.2007.12707.x},
    eprint = {https://academic.oup.com/mnras/article-pdf/382/4/1859/3958672/mnras0382-1859.pdf},
}

@article{PhysRevD.78.123520,
  title = {A hemispherical power asymmetry from inflation},
  author = {Erickcek, Adrienne L. and Kamionkowski, Marc and Carroll, Sean M.},
  journal = {Phys. Rev. D},
  volume = {78},
  issue = {12},
  pages = {123520},
  numpages = {5},
  year = {2008},
  month = {Dec},
  publisher = {American Physical Society},
  doi = {10.1103/PhysRevD.78.123520},
  url = {https://link.aps.org/doi/10.1103/PhysRevD.78.123520}
}

@article{Lew_2008,
doi = {10.1088/1475-7516/2008/09/023},
url = {https://doi.org/10.1088/1475-7516/2008/09/023},
year = {2008},
month = {sep},
publisher = {},
volume = {2008},
number = {09},
pages = {023},
author = {Lew, Bartosz},
title = {Hemispherical power asymmetry: parameter estimation from cosmic microwave background
WMAP5 data},
journal = {Journal of Cosmology and Astroparticle Physics}
}

@article{PhysRevD.78.103504,
  title = {Needlet detection of features in the WMAP CMB sky and the impact on anisotropies and hemispherical asymmetries},
  author = {Pietrobon, Davide and Amblard, Alexandre and Balbi, Amedeo and Cabella, Paolo and Cooray, Asantha and Marinucci, Domenico},
  journal = {Phys. Rev. D},
  volume = {78},
  issue = {10},
  pages = {103504},
  numpages = {12},
  year = {2008},
  month = {Nov},
  publisher = {American Physical Society},
  doi = {10.1103/PhysRevD.78.103504},
  url = {https://link.aps.org/doi/10.1103/PhysRevD.78.103504}
}

@article{Groeneboom2010ImprintsOA,
  title={Imprints of a hemispherical power asymmetry in the seven-year WMAP data due to non-commutativity of space-time},
  author={Nicolaas E. Groeneboom and Magnus Axelsson and David Fonseca Mota and Tomi S. Koivisto},
  journal={arXiv: Cosmology and Nongalactic Astrophysics},
  year={2010},
  url={https://api.semanticscholar.org/CorpusID:118782332}
}

@article{10.1093/mnras/stt1219,
    author = {Paci, F. and Gruppuso, A. and Finelli, F. and De Rosa, A. and Mandolesi, N. and Natoli, P.},
    title = {Hemispherical power asymmetries in the WMAP 7-year low-resolution temperature and polarization maps},
    journal = {Monthly Notices of the Royal Astronomical Society},
    volume = {434},
    number = {4},
    pages = {3071-3077},
    year = {2013},
    month = {07},
    issn = {0035-8711},
    doi = {10.1093/mnras/stt1219},
    url = {https://doi.org/10.1093/mnras/stt1219},
    eprint = {https://academic.oup.com/mnras/article-pdf/434/4/3071/18497532/stt1219.pdf},
}

@article{Notari_2014,
doi = {10.1088/1475-7516/2014/03/019},
url = {https://doi.org/10.1088/1475-7516/2014/03/019},
year = {2014},
month = {mar},
publisher = {},
volume = {2014},
number = {03},
pages = {019},
author = {Notari, Alessio and Quartin, Miguel and Catena, Riccardo},
title = {CMB aberration and Doppler effects as a source of hemispherical asymmetries},
journal = {Journal of Cosmology and Astroparticle Physics}
}

@article{JohnMcDonald_2013,
doi = {10.1088/1475-7516/2013/07/043},
url = {https://doi.org/10.1088/1475-7516/2013/07/043},
year = {2013},
month = {jul},
publisher = {},
volume = {2013},
number = {07},
pages = {043},
author = {John McDonald},
title = {Isocurvature and curvaton perturbations with red power spectrum and large hemispherical asymmetry},
journal = {Journal of Cosmology and Astroparticle Physics}
}

@article{PhysRevD.88.083527,
  title = {Hemispherical asymmetry and local non-Gaussianity: A consistency condition},
  author = {Namjoo, Mohammad Hossein and Baghram, Shant and Firouzjahi, Hassan},
  journal = {Phys. Rev. D},
  volume = {88},
  issue = {8},
  pages = {083527},
  numpages = {10},
  year = {2013},
  month = {Oct},
  publisher = {American Physical Society},
  doi = {10.1103/PhysRevD.88.083527},
  url = {https://link.aps.org/doi/10.1103/PhysRevD.88.083527}
}

@article{John_McDonald_2013,
doi = {10.1088/1475-7516/2013/11/041},
url = {https://doi.org/10.1088/1475-7516/2013/11/041},
year = {2013},
month = {nov},
publisher = {},
volume = {2013},
number = {11},
pages = {041},
author = {John McDonald},
title = {Hemispherical power asymmetry from scale-dependent modulated reheating},
journal = {Journal of Cosmology and Astroparticle Physics}
}

@article{PhysRevD.89.127303,
  title = {Hemispherical power asymmetry from a space-dependent component of the adiabatic power spectrum},
  author = {McDonald, John},
  journal = {Phys. Rev. D},
  volume = {89},
  issue = {12},
  pages = {127303},
  numpages = {5},
  year = {2014},
  month = {Jun},
  publisher = {American Physical Society},
  doi = {10.1103/PhysRevD.89.127303},
  url = {https://link.aps.org/doi/10.1103/PhysRevD.89.127303}
}

@article{McDonald_2014,
doi = {10.1088/1475-7516/2014/11/012},
url = {https://doi.org/10.1088/1475-7516/2014/11/012},
year = {2014},
month = {nov},
publisher = {},
volume = {2014},
number = {11},
pages = {012},
author = {McDonald, John},
title = {Negative running of the spectral index, hemispherical asymmetry and the consistency of Planck with large r},
journal = {Journal of Cosmology and Astroparticle Physics}
}

@article{10.1093/mnras/stu921,
    author = {Chluba, J. and Dai, Liang and Jeong, Donghui and Kamionkowski, Marc and Yoho, Amanda},
    title = {Linking the BICEP2 result and the hemispherical power asymmetry through spatial variation of r},
    journal = {Monthly Notices of the Royal Astronomical Society},
    volume = {442},
    number = {1},
    pages = {670-673},
    year = {2014},
    month = {06},
    issn = {0035-8711},
    doi = {10.1093/mnras/stu921},
    url = {https://doi.org/10.1093/mnras/stu921},
    eprint = {https://academic.oup.com/mnras/article-pdf/442/1/670/12653571/stu921.pdf},
}

@article{Firouzjahi_2014,
doi = {10.1088/1475-7516/2014/11/037},
url = {https://doi.org/10.1088/1475-7516/2014/11/037},
year = {2014},
month = {nov},
publisher = {},
volume = {2014},
number = {11},
pages = {037},
author = {Firouzjahi, Hassan and Gong, Jinn-Ouk and Namjoo, Mohammad Hossein},
title = {Scale-dependent hemispherical asymmetry from general initial state during inflation},
journal = {Journal of Cosmology and Astroparticle Physics}
}

@article{Namjoo_2014,
doi = {10.1088/1475-7516/2014/08/002},
url = {https://doi.org/10.1088/1475-7516/2014/08/002},
year = {2014},
month = {aug},
publisher = {},
volume = {2014},
number = {08},
pages = {002},
author = {Namjoo, Mohammad Hossein and Abolhasani, Ali Akbar and Baghram, Shant and Firouzjahi, Hassan},
title = {CMB hemispherical asymmetry: long mode modulation and non-Gaussianity},
journal = {Journal of Cosmology and Astroparticle Physics}
}

@article{Assadullahi_2015,
doi = {10.1088/1475-7516/2015/04/017},
url = {https://doi.org/10.1088/1475-7516/2015/04/017},
year = {2015},
month = {apr},
publisher = {},
volume = {2015},
number = {04},
pages = {017},
author = {Assadullahi, Hooshyar and Firouzjahi, Hassan and Namjoo, Mohammad Hossein and Wands, David},
title = {CMB hemispherical asymmetry from non-linear isocurvature perturbations},
journal = {Journal of Cosmology and Astroparticle Physics}
}

@article{PhysRevD.91.062002,
  title = {Hemispherical asymmetry from an isotropy violating stochastic gravitational wave background},
  author = {Mukherjee, Suvodip},
  journal = {Phys. Rev. D},
  volume = {91},
  issue = {6},
  pages = {062002},
  numpages = {6},
  year = {2015},
  month = {Mar},
  publisher = {American Physical Society},
  doi = {10.1103/PhysRevD.91.062002},
  url = {https://link.aps.org/doi/10.1103/PhysRevD.91.062002}
}

@misc{mukherjee2016unifiedoriginhemisphericalasymmetry,
      title={Unified origin of hemispherical asymmetry in scalar and tensor perturbations}, 
      author={Suvodip Mukherjee and Tarun Souradeep},
      year={2016},
      eprint={1504.02285},
      archivePrefix={arXiv},
      primaryClass={astro-ph.CO},
      url={https://arxiv.org/abs/1504.02285}, 
}

@article{PhysRevLett.116.221301,
  title = {Litmus Test for Cosmic Hemispherical Asymmetry in the Cosmic Microwave Background $B$-Mode Polarization},
  author = {Mukherjee, Suvodip and Souradeep, Tarun},
  journal = {Phys. Rev. Lett.},
  volume = {116},
  issue = {22},
  pages = {221301},
  numpages = {6},
  year = {2016},
  month = {Jun},
  publisher = {American Physical Society},
  doi = {10.1103/PhysRevLett.116.221301},
  url = {https://link.aps.org/doi/10.1103/PhysRevLett.116.221301}
}

@article{Mukherjee_2016,
doi = {10.1088/1475-7516/2016/06/042},
url = {https://doi.org/10.1088/1475-7516/2016/06/042},
year = {2016},
month = {jun},
publisher = {},
volume = {2016},
number = {06},
pages = {042},
author = {Mukherjee, Suvodip and Aluri, Pavan K. and Das, Santanu and Shaikh, Shabbir and Souradeep, Tarun},
title = {Direction dependence of cosmological parameters due to  cosmic hemispherical  asymmetry},
journal = {Journal of Cosmology and Astroparticle Physics}
}

@article{Byrnes_2016,
doi = {10.1088/1475-7516/2016/06/025},
url = {https://doi.org/10.1088/1475-7516/2016/06/025},
year = {2016},
month = {jun},
publisher = {},
volume = {2016},
number = {06},
pages = {025},
author = {Byrnes, Christian T. and Regan, Donough and Seery, David and Tarrant, Ewan R.M.},
title = {The hemispherical asymmetry from a scale-dependent inflationary bispectrum},
journal = {Journal of Cosmology and Astroparticle Physics}
}

@article{PhysRevD.96.083516,
  title = {Hemispherical power asymmetry of the cosmic microwave background from a remnant of a pre-inflationary topological defect},
  author = {Yang, Qiaoli and Liu, Yunqi and Di, Haoran},
  journal = {Phys. Rev. D},
  volume = {96},
  issue = {8},
  pages = {083516},
  numpages = {7},
  year = {2017},
  month = {Oct},
  publisher = {American Physical Society},
  doi = {10.1103/PhysRevD.96.083516},
  url = {https://link.aps.org/doi/10.1103/PhysRevD.96.083516}
}

@article{Jazayeri_2014,
doi = {10.1088/1475-7516/2014/11/044},
url = {https://doi.org/10.1088/1475-7516/2014/11/044},
year = {2014},
month = {nov},
publisher = {},
volume = {2014},
number = {11},
pages = {044},
author = {Jazayeri, Sadra and Akrami, Yashar and Firouzjahi, Hassan and Solomon, Adam R. and Wang, Yi},
title = {Inflationary power asymmetry from primordial domain walls},
journal = {Journal of Cosmology and Astroparticle Physics}
}

@article{Quartin_2015,
doi = {10.1088/1475-7516/2015/01/008},
url = {https://doi.org/10.1088/1475-7516/2015/01/008},
year = {2015},
month = {jan},
publisher = {},
volume = {2015},
number = {01},
pages = {008},
author = {Quartin, Miguel and Notari, Alessio},
title = {On the significance of power asymmetries in Planck CMB data at all scales},
journal = {Journal of Cosmology and Astroparticle Physics}
}

@article{Byrnes_2015,
doi = {10.1088/1475-7516/2015/07/007},
url = {https://doi.org/10.1088/1475-7516/2015/07/007},
year = {2015},
month = {jul},
publisher = {},
volume = {2015},
number = {07},
pages = {007},
author = {Byrnes, Christian T. and Tarrant, Ewan R.M.},
title = {Scale-dependent non-Gaussianity and the CMB power asymmetry},
journal = {Journal of Cosmology and Astroparticle Physics}
}

@article{PhysRevD.92.023505,
  title = {Generating the cosmic microwave background power asymmetry with ${g}_{NL}$},
  author = {Kenton, Zachary and Mulryne, David J. and Thomas, Steven},
  journal = {Phys. Rev. D},
  volume = {92},
  issue = {2},
  pages = {023505},
  numpages = {6},
  year = {2015},
  month = {Jul},
  publisher = {American Physical Society},
  doi = {10.1103/PhysRevD.92.023505},
  url = {https://link.aps.org/doi/10.1103/PhysRevD.92.023505}
}

@article{PhysRevD.92.064038,
  title = {Loop quantum cosmology, non-Gaussianity, and CMB power asymmetry},
  author = {Agullo, Ivan},
  journal = {Phys. Rev. D},
  volume = {92},
  issue = {6},
  pages = {064038},
  numpages = {6},
  year = {2015},
  month = {Sep},
  publisher = {American Physical Society},
  doi = {10.1103/PhysRevD.92.064038},
  url = {https://link.aps.org/doi/10.1103/PhysRevD.92.064038}
}

@article{Wang_2016,
doi = {10.1088/1475-7516/2016/02/019},
url = {https://doi.org/10.1088/1475-7516/2016/02/019},
year = {2016},
month = {feb},
publisher = {},
volume = {2016},
number = {02},
pages = {019},
author = {Wang, Dong-Gang and Cai, Yi-Fu and Zhao, Wen and Zhang, Yang},
title = {Scale-dependent CMB power asymmetry from primordial speed of sound and a generalized δ N formalism},
journal = {Journal of Cosmology and Astroparticle Physics}
}

@article{PhysRevD.93.123003,
  title = {Implications of the cosmic microwave background power asymmetry for the early universe},
  author = {Byrnes, Christian T. and Regan, Donough and Seery, David and Tarrant, Ewan R. M.},
  journal = {Phys. Rev. D},
  volume = {93},
  issue = {12},
  pages = {123003},
  numpages = {5},
  year = {2016},
  month = {Jun},
  publisher = {American Physical Society},
  doi = {10.1103/PhysRevD.93.123003},
  url = {https://link.aps.org/doi/10.1103/PhysRevD.93.123003}
}

@article{PhysRevD.97.043501,
  title = {Primordial non-Gaussianity and power asymmetry with quantum gravitational effects in loop quantum cosmology},
  author = {Zhu, Tao and Wang, Anzhong and Kirsten, Klaus and Cleaver, Gerald and Sheng, Qin},
  journal = {Phys. Rev. D},
  volume = {97},
  issue = {4},
  pages = {043501},
  numpages = {11},
  year = {2018},
  month = {Feb},
  publisher = {American Physical Society},
  doi = {10.1103/PhysRevD.97.043501},
  url = {https://link.aps.org/doi/10.1103/PhysRevD.97.043501}
}

@article{PhysRevD.97.063504,
  title = {Closing in on the large-scale CMB power asymmetry},
  author = {Contreras, D. and Hutchinson, J. and Moss, A. and Scott, D. and Zibin, J. P.},
  journal = {Phys. Rev. D},
  volume = {97},
  issue = {6},
  pages = {063504},
  numpages = {6},
  year = {2018},
  month = {Mar},
  publisher = {American Physical Society},
  doi = {10.1103/PhysRevD.97.063504},
  url = {https://link.aps.org/doi/10.1103/PhysRevD.97.063504}
}
